\documentclass[preprint,3p]{elsarticle}
\usepackage{color}
\usepackage{amssymb}
\usepackage{mathrsfs}
\usepackage{epstopdf}
\usepackage{makeidx}
\usepackage{bm}
\usepackage{psfrag}
\usepackage{amsthm}
\usepackage{algorithm}
\usepackage{algorithmic}
\usepackage{epsf,amsfonts,amsmath}
\usepackage{amssymb,amsmath}
\usepackage{appendix}
\usepackage{graphicx}
\usepackage{multirow}
\usepackage{array}

\newtheorem{remark1}{Remark}

\newtheorem{remark4}{Corollary}

\newtheorem{theorem}{Theorem}
\usepackage{caption}
\usepackage{times}
\usepackage{float}
\usepackage[caption = false]{subfig}
\usepackage[intoc]{nomencl}
\makenomenclature
\renewcommand\nomgroup[1]{%
\ifthenelse{\equal{#1}{}}{%
\item[\textbf{Symbols}] }{
\ifthenelse{\equal{#1}{R}}{%
\item[\textbf{Roman Symbols}]}{
\ifthenelse{\equal{#1}{G}}{%
\item[\textbf{Greek Symbols }]}{
\ifthenelse{\equal{#1}{S}}{%
\item[\textbf{Superscripts  }]}{{
\ifthenelse{\equal{#1}{U}}{%
\item[\textbf{Subscripts }]}{{
\ifthenelse{\equal{#1}{X}}{%
\item[\textbf{Other Symbols }]}
{{}}}}}}}}}}
\newcommand{\argmin}{\arg\!\min}
\newcommand\floor[1]{\lfloor#1\rfloor}
\journal{Elsevier}

\begin{document}

\begin{frontmatter}

\title{Least Squares Polynomial Chaos Expansion: A Review of Sampling Strategies}

\author[UCB1]{Mohammad Hadigol}
\ead{mohammad.hadigol@colorado.edu}

\author[UCB1]{Alireza Doostan\corref{cor1}}
\ead{alireza.doostan@colorado.edu}

\cortext[cor1]{Corresponding Author: Alireza Doostan}

\address[UCB1]{Smead Aerospace Engineering Sciences Department, University of Colorado, Boulder, CO 80309, USA}

\begin{abstract}
\label{Abstract}
{\color{black}
As non-institutive polynomial chaos expansion (PCE) techniques have gained growing popularity among researchers, we here provide a comprehensive review of major sampling strategies for the least squares based PCE. Traditional sampling methods, such as Monte Carlo, Latin hypercube, quasi-Monte Carlo, optimal design of experiments (ODE), Gaussian quadratures, as well as more recent techniques, such as coherence-optimal and randomized quadratures are discussed. We also propose a hybrid sampling method, dubbed {\it alphabetic-coherence-optimal}, that employs the so-called alphabetic optimality criteria used in the context of ODE in conjunction with coherence-optimal samples. A comparison between the empirical performance of the selected sampling methods applied to three numerical examples, including high-order PCE's, high-dimensional problems, and low oversampling ratios, is presented to provide a road map for practitioners seeking the most suitable sampling technique for a problem at hand. We observed that the  alphabetic-coherence-optimal technique outperforms other sampling methods, specially when high-order ODE are employed and/or the oversampling ratio is low. }
\end{abstract}


\end{frontmatter}

%

%
\section{Introduction}
\label{sec:introduction}

Reliable and rigorous simulations of real world engineering problems involve characterization of the often uncertain system inputs, e.g., material properties, initial, or boundary conditions, and quantification of their impact on the output quantities of interest (QoI's). This is the subject of uncertainty quantification (UQ), an emerging field in computational engineering and science, which aims at providing tools for assessing the credibility of model predictions and facilitating decision making under uncertainty.  

A major class of UQ techniques are probabilistic where uncertain parameters are represented by random variables or processes. Among the probabilistic UQ approaches, stochastic spectral methods based on polynomial chaos (PC) expansions \cite{Wiener38,Cameron47,Ghanem91a,Xiu10a} have received special attention due to their advantages over traditional UQ techniques; see, e.g., \cite{Ghanem91a,Sudret08d,Najm09,Doostan11a, hadigol15}. Let $\left(\Omega,\mathcal{T},\mathcal{P} \right)$ be a complete probability space where $\Omega$ is the sample set (or design space in the context of design of experiments) and $\mathcal{P}$ is a probability measure on the $\sigma-$field $\mathcal{T}$. Also assume that the system input uncertainty is described by the random vector $\bm{\Xi} = \left(\Xi_1,\cdots, \Xi_d\right):\Omega \rightarrow \mathbb{R}^{d}$, $d \in \mathbb{N}$, consisting of $d$ independent identically distributed (i.i.d.) random variables $\Xi_k$ with realizations $\bm\xi = (\xi_1,\cdots,\xi_d)$. The marginal probability density function (pdf) of  $\Xi_k$ and the joint pdf of $\bm{\Xi}$ are denoted by $f(\xi_k)$ and $f(\bm{\xi})=\prod_{k=1}^d f(\xi_k)$, respectively. The PC representation of $u(\bm{\Xi})$, a scalar QoI with finite variance, is given by 
\begin{equation} 
\label{eqn:PC_expansion_inf}
{u}(\bm{\Xi}) = \sum_{j=1}^{\infty} c_j \psi_{j}(\bm{\Xi}),
\end{equation}
where $\{\psi_{j}(\bm{\Xi})\}$ is a set of multivariate polynomials orthonormal with respect to $f(\bm{\xi})$ and $c_{j}$ are the deterministic PC coefficients to be determined. For computational purpose, the infinite series in (\ref{eqn:PC_expansion_inf}) may be truncated by retaining the first $P$ terms as
\begin{equation} 
\label{eqn:PC_expansion_P}
{u}(\bm{\Xi}) = \sum_{j=1}^{P} c_j \psi_{j}(\bm{\Xi}) + \epsilon(\bm{\Xi}),  
\end{equation}
where $\epsilon(\bm{\Xi})$ represents the random truncation error. A detailed description of the construction of the PC basis $\{\psi_{j}(\bm{\Xi})\}$ and the truncated expansion in (\ref{eqn:PC_expansion_P}) will be discussed in Section \ref{sec:Polynomial_chaos_expansion}. 

The main task in PC-based methods is to compute the PC coefficients either intrusively or non-intrusively \citep{Ghanem91a,LeMaitre10}. In an intrusive approach, {\color{black}the governing equations are} projected onto the subspace spanned by the PC basis via the Galerkin formulation \citep{Ghanem91a}, often requiring some modifications to the existing deterministic solvers that may not be desirable. On the other hand, non-intrusive methods are based on sampling and treat existing deterministic solvers as a black box. Sampling-based techniques such as least squares approximation (LSA) \citep{Hosder06,Hampton15b}, pseudo-spectral collocation \citep{LeMaitre10,Constantine12a}, Monte Carlo sampling \citep{LeMaitre10}, and compressive sampling \citep{Doostan11a, Hampton16Chapter} have been employed in the literature to identify the PC coefficients. 

In this work, we consider the LSA approach to determine the PC coefficients. From the statistical point of view, LSA is similar to the regression of the exact solution $u(\bm{\Xi})$ in the PC basis \citep{Hosder06}. Let us denote the $i$th realization of $\bm \Xi$ by $\bm{\xi}^{(i)}$, a.k.a a design point, generated via random sampling, for instance, according to $f(\bm\xi)$, and the corresponding realization of ${u}(\bm{\Xi})$ by ${u}(\bm{\xi}^{(i)})$. Given the pair $\{ \bm{\xi}^{(i)} \}_{i=1}^{N}$ and $\{ {u}(\bm{\xi}^{(i)}) \}_{i=1}^{N}$, with $N$ being the number of independent samples, the discrete representation of (\ref{eqn:PC_expansion_P}) can be written as
\begin{equation}
\label{eqn:regression5}
\bm{u} = \bm{\Psi} \bm{c} + \bm{\epsilon},
\end{equation}
where $\bm{u} = ( u({\bm{\xi}}^{(1)}), \cdots,  u({\bm{\xi}}^{(N)}) )^T \in \mathbb{R}^N$ contains the realizations of the QoI $u$, $\bm{\Psi}(i,j)=\psi_j({\bm{\xi}}^{(i)}) \in \mathbb{R}^{N \times P}$ is the measurement matrix containing samples of the PC basis, $\bm{c} = (c_1, \cdots,  c_P )^T \in \mathbb{R}^P$ is the vector of unknown PC coefficients, a.k.a. the estimators, and the vector $\bm{\epsilon} \in \mathbb{R}^N$ contains unknown truncation errors between the solution realizations $u$ and the truncated PC approximations of $u$. A diagonal positive-definite matrix $\bm W$ is also introduced such that $\bm W(i,i)=w(\bm{\xi}^{(i)})$ is a function of the sample points, which in turn depends on the sampling strategy, and will be discussed later in Section \ref{sec:Sampling techniques}. The unknown coefficients $\bm{c}$ may be approximated by solving the least squares problem
\begin{equation}
\label{eqn:regression1}
\min_{\color{black}{\bm{c}}} \Vert \bm{W} \bm{u} - \bm{W} \bm{\Psi} \bm{c} \Vert_2,
\end{equation}
where $\Vert\cdot\Vert_2$ is the standard Euclidean norm. The solution to (\ref{eqn:regression1}) {\color{black}is} computed from the normal equation 
\begin{equation}
\label{eqn:regression2}
(\bm{W}  \bm{\Psi})^T (\bm{W} \bm{\Psi}) \hat{\bm{c}} = (\bm{W} \bm{\Psi})^T \bm{W} \bm{u}.
\end{equation}

The computational cost of constructing the PCE via (\ref{eqn:regression1}) is controlled by the number of samples $N$ and may be high when evaluating the QoI is expensive. In general, a stable solution $\bm c$ to (\ref{eqn:regression1}) requires $N \geq P$ (often $N\gg P$) samples of $u$ \citep{Hampton15b} and, as we shall see in Section \ref{sec:Polynomial_chaos_expansion}, $P$ grows quickly when either the dimensionality of the problem, $d$, or the total order of the PCE, $p$, is increased. Consequently, accurate approximation of $\bm c$ with as small $N$ as possible becomes a critical task. This has motivated researchers to explore various strategies for constructing the sample set $\{ \bm{\xi}^{(i)} \}_{i=1}^{N}$ beyond the standard MC sampling according to $f(\bm\xi)$. In general, one may categorize the sampling techniques for the LSA problem into two groups: random and deterministic. Unlike random sampling, deterministic sampling techniques attempt to select the sample points $\{ \bm{\xi}^{(i)} \}_{i=1}^{N}$ in a deterministic fashion.

Among the early work on sampling techniques for PCE via LSA, we mention \cite{Choi_2004} in which Latin hypercube (LH) sampling is used instead of the standard MC sampling. Recently, an importance sampling technique referred to as {\it asymptotic} was studied in \citep{Hampton15b,Feldhacker16,Narayan_2014} based on asymptotic (in order) analysis of $d$-dimensional Legendre and Hermite polynomials. This approach leads to sampling from a $d$-dimensional Chebyshev distribution for the case of Legendre polynomials and sampling uniformly from a $d$-dimensional ball (with a radius determined by the total order of the PC expansion) for Hermite polynomials. Additionally, for arbitrary dimension and order, a different importance sampling approach dubbed {\it coherence-optimal} was proposed in \citep{Hampton15b} which ensures an accurate computation of $\bm c$ with a number of solution realizations that depends linearly on $P$ (up to a logarithmic factor). A related sampling method has also been presented in \citep{Cohen16}. Random sampling from tensor product of Gaussian quadrature points has been also proposed in \cite{Zhou_2015}, where the authors prove an asymptotic (in order) stability of this method with a number of points which scales linearly (up to a logarithmic factor) with $P$. 

A quasi-Monte Carlo (QMC) method using the so-called low-discrepancy points was employed in \cite{Migliorati15} to deterministically select the sample points for the least squares PCE such that a stable and accurate solution is achieved with a number of samples $N$ proportional to the square of the size of PC basis, $P$, up to logarithmic factors. Another deterministic sampling approach based on Weil's theorem was proposed in \cite{Zhou_2014} which leads to a stable approximation in the Chebyshev basis with a number of samples that scales quadratically in $P$. 

Optimal sampling of the design points $\{ \bm{\xi}^{(i)} \}_{i=1}^{N}$ for general regression models has received extensive attention in statistics community \citep{Fedorov72, Fedorov1997, Box2005, Pukelsheim2006, Atkinson08} and is referred to as the optimal design of experiments (ODE). The alphabetic optimality criteria for ODE, a.k.a. \textit{classical optimality}, either focus on minimizing the {\it error} in estimating the unknown coefficients, or {\it parameters}, $\bm c$, e.g., \textit{D}-, \textit{A}-, and \textit{E}-optimal designs, or the error in the prediction of $u$, e.g., \textit{I}-optimal design. As explained in Section~\ref{sec:optimal ODE}, these optimality criterion are mostly based on some functional of the information matrix $\bm{M}$ given by
\begin{equation}
\label{eqn:info_matrix}
\bm{M}=\frac{1}{N}\bm{\Psi}^{T} \bm{\Psi},
\end{equation}
when the standard LSA is considered, and 
\begin{equation}
\label{eqn:info_matrix_weighted}
\bm{M}=\frac{1}{N}(\bm{W}\bm{\Psi})^{T} (\bm{W}\bm{\Psi}),
\end{equation}
for the case of weighted LSA problem (\ref{eqn:regression1}). For example, \textit{D}-optimal criterion takes the determinant of $\bm{M}^{-1}$, denoted by $|\bm{M}^{-1}|$, and constructs $\{ \bm{\xi}^{(i)} \}_{i=1}^{N}$ such that $|\bm{M}^{-1}|$ is minimized. Among the limited work utilizing ODE for PCE, we mention \citep{Zein2013} in which \textit{D}-optimality criterion was used to generate the sample set. It was shown that a Fedorov-genetic algorithm used to generate the \textit{D}-optimal sample set could compute the PC coefficients accurately with a smaller cost in comparison to other common sampling techniques. In \cite{Burnaev2016}, $D$-optimal criterion has been employed to efficiently estimate the PCE-based Sobol' indices via the LSA approach. Very recently, a quasi-optimal sampling approach for the weighted LSA problem has been proposed in \cite{Shin_2016} which takes a function that involves $|\bm{M}|$ as the optimality measure.  

Given the growing interest in designing optimal sampling strategies for PCE's, the aims of this manuscript are three fold. Firstly, we provide a comprehensive review of both standard and advanced (random and deterministic) techniques for choosing samples $\{ \bm{\xi}^{(i)} \}_{i=1}^{N}$. Secondly, we propose the so-called {\it alphabetic-coherence-optimal} sampling technique, a hybrid scheme which combines the coherence-optimal approach of \citep{Hampton15b} with classical alphabetic optimality criteria. Thirdly, we provide a comparison study of the performance of various sampling strategies on three example problems. Our ultimate goal is to provide practitioners with a collection of available options and a set of comparative studies as how these various sampling techniques perform. 

The remainder of this paper is organized as follows. Section \ref{sec:Method synopsis} summarizes the problem formulation, where we start with an abstract problem and provide details on the construction of PC basis. Following that, in Section \ref{sec:Definitions and background}, some definitions and preliminaries used in describing the sampling techniques are presented. Standard MC, asymptotic, coherence-optimal and random quadrature sampling techniques, as well as space-filling methods, such as quasi-MC and LH sampling, are reviewed in Section \ref{sec:Sampling techniques}. Moreover, Section \ref{sec:Sampling techniques} provides a comprehensive review on the application of various alphabetic optimality criteria in the context of least squares PCE's. Finally, three numerical examples concerning the recovery of randomly generated functions, a nonlinear stochastic Duffing oscillator, and the estimation of the remaining useful life of Lithium-ion batteries are considered in Section \ref{sec:Numerical examples} to study the empirical performance of a number of sampling methods.   

\section{Problem statement and polynomial chaos expansion}
\label{sec:Method synopsis}

\subsection{Problem statement}
\label{sec:Problem statement}

In the probabilistic framework, model uncertainties are represented using random variables and processes, leading to non-deterministic governing equations, 
\begin{equation}
\label{eqn:full_SPDE}
\mathcal{R}\left(\bm{x},t,\bm{\Xi};u\right)=\bm 0,\quad (\bm{x},t,\bm \Xi)\in \mathcal{D}\times[0,T]\times\Omega.
\end{equation}
Here, $\mathcal{R}$ denotes the set of linear or non-linear governing equations, $\bm{x}$ the spatial variable, $t$ time, $\bm{\Xi}$ the set of $d$ independent random variables characterizing model uncertainties, as introduced in Section~\ref{sec:introduction}, and $u$ the solution. One of the goals in uncertainty propagation is to construct an approximation of the mapping $\bm \Xi\rightarrow u(\bm x,t,\bm \Xi)$, here, using a set of samples $\{ \bm{\xi}^{(i)} \}_{i=1}^{N}$ and the corresponding solution realizations $\{ u(\bm x,t,\bm{\xi}^{(i)})\}_{i=1}^{N}$, or samples of a QoI depending on $u$. For the interest of a simpler presentation, we assume such a construction is desired for a fixed spatial location $\bm x$ and time instance $t$, and hereafter suppress the dependence of $u$ on $\bm x$ and $t$.

\subsection{Polynomial chaos expansion}
\label{sec:Polynomial_chaos_expansion}

PCE provides a framework for approximating the solution $u$ in (\ref{eqn:full_SPDE}) of a stochastic system, by projecting it onto a basis of polynomials of the random inputs. It was first proposed by Wiener \citep{Wiener38} and introduced to the engineering field by Ghanem and Spanos \citep{Ghanem91a} for problems with Gaussian random inputs, and later extended {\color{black}to} non-Gaussian random inputs by using polynomials of the Askey scheme (generalized PCE) \citep{Xiu02}.  The PC representation of the finite variance QoI, $u$, is given in (\ref{eqn:PC_expansion_inf}) with the associated truncated expansion in (\ref{eqn:PC_expansion_P}). To specify the truncation, we first describe the construction of the multi-dimensional basis functions $\psi_j(\bm\Xi)$ adopted here. Let $\{\psi_{j_k}(\Xi_k):\ j_k \in \mathbb{N}_0\}$,  with $\mathbb{N}_0 := \mathbb{N}\cup \lbrace 0\rbrace$, denote a complete set of univariate polynomials of order (degree) $j_k$ orthonormal with respect to $f(\xi_k)$; See Table \ref{table:Askey}. The multi-dimensional basis functions are (tensor) products of the univariate functions, i.e., 
\begin{equation} 
\label{eqn:multivariate_polynomials}
\psi_{\bm{j}}(\bm{\Xi}) = \prod_{k=1}^{d} \psi_{{j}_k}(\Xi_k),
\end{equation}
where the multi-index $\bm j=(j_1,\cdots,j_d)$ specifies the order of $\psi_{\bm{j}}(\bm{\Xi})$ in each of the dimensions $\Xi_k$. For convenience, in (\ref{eqn:PC_expansion_inf}) we have considered an integer indexing of each $\psi_{\bm{j}}(\bm{\Xi})$ so that there is a one-to-one correspondence between the elements of $\{\psi_{\bm{j}}(\bm{\Xi})\}$ and $\{\psi_j(\bm{\Xi})\}$. For computation, in (\ref{eqn:PC_expansion_P}), we consider a truncation of (\ref{eqn:PC_expansion_inf}) where we limit the total order of $\{\psi_{\bm{j}}(\bm{\Xi})\}$ to $p$; that is, we consider only multi-indices ${\bm{j}}$ for which $\sum_{k=1}^d j_k\le p$. With this truncation, the number $P$ of basis functions retained in (\ref{eqn:PC_expansion_P}) is given by the factorial relation
\begin{equation} 
\label{eqn:P+1}
P= \frac{(p+d)!}{p!d!}.
\end{equation}
The truncated PC expansion in (\ref{eqn:PC_expansion_P}) converges in the mean-square sense as $ p \ \ (\text{hence} \ \ P) \rightarrow \infty$, when $u$ has finite variance and the coefficients $c_{\bm{i}}$ are computed from the projection equation
\begin{equation} 
\label{eq:projection}
c_j = \mathbb{E}[u \psi_j].
\end{equation}
In practice, it may not be feasible or efficient to compute $c_j$ directly from (\ref{eq:projection}) and alternative approaches, such as LSA in (\ref{eqn:regression1}), may be preferred.

\begin{table}[h]
\small
\caption{Correspondence of Wiener-Askey polynomial chaos and {\color{black}probability} distribution of the random variables \citep{Xiu02}.} 
\centering
\begin{tabular}[t]{ c   c  c }   
\hline \hline
{\color{black}$\Xi_k$} & {\color{black}Polynomial type} & Support \\  
\hline \hline
Gaussian & Hermite & (-$\infty$,+$\infty$) \\  
\hline
Gamma & Laguerre & (0,+$\infty$) \\  
\hline
Beta & Jacobi & [a,b] \\  
\hline
Uniform & Legendre & [a,b] \\  
\hline
\end{tabular}
\label{table:Askey} 
\end{table}

\section{Background on stability and convergence of least squares polynomial chaos expansion} 
\label{sec:Definitions and background}

In order to keep this paper self-contained, we next present a brief summary of results on stability and convergence of the solution to least squares problem (\ref{eqn:regression1}), highlighting the role of some key quantities.

\subsection{Stability of least squares approximation}
\label{sec:Stability}
One of sampling design goals is to improve the {\it stability} of (\ref{eqn:regression1}), in that small perturbations in the samples of $u$ do not lead to large variations in the computed $\bm c$ and, consequently, the PC approximation of $u$. Following \cite{Cohen2013,Hampton15b}, a key stability quantity is how close $\bm{M}$ is to the identity, given by $\Vert \bm{M}-\bm I\Vert$, where $\Vert \cdot\Vert$ denotes the spectral norm of a matrix. Due to the orthonormality of the PC basis, $\bm{M}$ is on average identity; however, it deviates from the identity for finite samples sizes $N$. A small $\Vert \bm{M}-\bm I\Vert$ ensures $\bm{M}$ is invertible and has small condition number, and that (\ref{eqn:regression2}) leads to a stable approximation of $\bm{c}$. These statements are quantified in the following corollary, reporting results from \cite{Rauhut10,Cohen2013,Hampton15b}.

\begin{remark4}
\label{cor:stability}
Let $\hat{u}(\bm\Xi)=\mathop{\sum}\limits_{j=1}^P\hat{c}_j\psi_j(\bm{\Xi})$. Let $\hat{\bm{c}}$ and $\hat{\bm{c}}_\delta$ be the solution to (\ref{eqn:regression2}) corresponding to data $\bm u$ and $\bm u+\delta\bm u$, respectively, where $\delta\bm u$ is a perturbation in $\bm u$. Let $\Vert \bm{M}-\bm I\Vert\le \delta$ for some $0\le \delta <1$. Then, 
\begin{align}
&(1-\delta)\le \lambda_{\min}(\bm{M})\le \lambda_{\max}(\bm{M})\le (1+\delta);\label{eqn:stab_bound1}\\
&\mathcal{K}(\bm{M})=\frac{ \lambda_{\max}(\bm{M})}{\lambda_{\min}(\bm{M})}\le \frac{1+\delta}{1-\delta};\label{eqn:stab_bound2}\\
&\Vert\hat{u}\Vert_{L_2(\Omega,f)}=\Vert\hat{\bm{c}}\Vert\le \frac{(1+\delta)^{1/2}}{1-\delta}\frac{1}{\sqrt{N}}\Vert \bm{Wu}\Vert \label{eqn:stab_bound3},
\end{align}
where $\mathcal{K}(\bm{M})$ is the condition number of $\bm{M}$. Additionally, 
\begin{align}
\label{eqn:perturb_bound}
\frac{\Vert \hat{\bm{c}}_\delta - \hat{\bm{c}}\Vert}{\Vert \hat{\bm{c}}\Vert}\le \mathcal{K}(\bm{M}) \frac{\Vert \bm W\delta\bm u\Vert}{\Vert \bm W \bm u\Vert}.
\end{align}
\end{remark4}
\begin{proof}
See \ref{apx:proof_cor} for the derivation of (\ref{eqn:stab_bound1})-(\ref{eqn:perturb_bound}).
\end{proof}

In particular, (\ref{eqn:stab_bound3}) and (\ref{eqn:perturb_bound}) indicate that small changes in $\bm u$ do not lead to large changes in $\hat{\bm c}$ -- hence the stability of the least squares problem -- as long as $\delta$ is small. As we shall explain in Section \ref{sec:Sampling techniques}, the results in Corollary \ref{cor:stability} form the basis of designing sampling strategies for PCE via LSA. 

\subsection{Coherence parameter definition}
\label{sec:Coherence-Parameter definition}

Let $w(\bm\xi)$ denote a weight function depending on the probability density function $f_{\bm Y}(\bm\xi)$ used to generate $\bm \xi$, possibly different from $f(\bm\xi)$, which we shall specify in Sections \ref{sec:Coherence-optimal sampling} and \ref{sec:Asymptotic sampling}. We note that, for the sake of a simpler notation, we refer to all realized input random vectors by $\bm \xi$, regardless of the sampling distribution used to generate these realizations. In \citep{Hampton15b}, the {\it coherence parameter} $\mu$ given by
\begin{equation} 
\label{eqn:coh-par}
\mu(\bm Y) := \sup_{\color{black}{\bm{\xi} \in \Omega}} \sum_{j=1}^P |w(\bm \xi) \psi_j(\bm \xi)|^2
\end{equation}
is used to bound (probabilistically) the deviation of $\bm M$ from the identity for stability purposes as discussed in Section \ref{sec:Stability}. In particular, \cite{Cohen2013} shows that for $0\le\delta<1$,
\begin{equation}
\label{eqn:prob_stability_bound}
\mathbb{P}\left(\Vert \bm M - \bm I\Vert\le \delta \right)\ge 1- 2 P\exp(-c_\delta N \mu^{-1}),
\end{equation}
where $c_\delta = \delta + (1-\delta)\log(1-\delta)$ and $\mathbb{P}(\cdot)$ denotes the probability of an event. Notice that  (\ref{eqn:prob_stability_bound}) suggests bounding $\Vert \bm M - \bm I\Vert$ with high probability requires a sampling rate that linearly depends on $\mu$. More precisely, the following theorem, adopted from \citep[Section 2.3]{Hampton15b}, demonstrates the dependence of sample size $N$ on the coherence parameter $\mu$ -- or indirectly the sampling distribution -- for the convergence of PCE with bounded polynomials, e.g., of Legendre type. 

\begin{theorem}[\citep{Hampton15b}]
\label{thm:SampleDepth}
Let $\hat{u}(\bm\Xi)=\mathop{\sum}\limits_{j=1}^P\hat{c}_j\psi_j(\bm{\Xi})$, where $\hat{\bm{c}}=(\hat{c}_1,\cdots,\hat{c}_P)^T$ is the solution to (\ref{eqn:regression2}).
For $\mathcal{E}$, a sampling event that occurs with probability
\begin{align*}
\mathbb{P}(\mathcal{E})\ge1 - 2P\exp(-0.1N\mu^{-1}),
\end{align*}
it follows that
\begin{align*}
\mathbb{E}\left(\|u-\hat{u}\|^2_{L_2(\Omega,f)};\mathcal{E}\right)\le \mathbb{E}(\epsilon^2)\left(1+\frac{4\mu}{N}\right),
\end{align*}
where
\begin{align}
\label{eqn:restricted_expectation}
\mathbb{E}(X;\mathcal{E}) = \int_{\mathcal{E}}X(\bm{\xi})f(\bm{\xi})d\bm{\xi} =\mathbb{E}(X|\mathcal{E})\mathbb{P}(\mathcal{E})
\end{align}
denotes the expectation restricted to the event $\mathcal{E}$.
\end{theorem}

When $\mu$ in (\ref{eqn:coh-par}) is infinite, i.e., for unbounded polynomials such as Hermite polynomials, a coherence parameter defined on an appropriate subset of $\Omega$ is instead considered. The interested reader is referred to \citep[Section 2.2]{Hampton15b} for further details. 

\section{Sampling techniques}
\label{sec:Sampling techniques}

In this section, we start reviewing the standard MC technique followed by the coherence-optimal and asymptotic sampling methods presented in Sections \ref{sec:Coherence-optimal sampling} and \ref{sec:Asymptotic sampling}, respectively. Similar to MC sampling, the last two sampling strategies are random. However, unlike standard MC sampling, they seek sampling distributions, possibly different from $f(\bm\xi)$, that lead to some optimality. We then turn our attention to methods that directly generate discrete sample points $\{ \bm{\xi}^{(i)} \}_{i=1}^{N}$, starting by stochastic collocation via deterministic and randomized quadrature points in Section \ref{sec:SC}. In Section \ref{sec:optimal ODE}, we provide a detailed description of the alphabetic optimal designs and present a sequential technique to construct these designs. Finally, space-filling designs, such as QMC and LH sampling techniques as two instances of widely-used, random sampling strategies in designing computer experiments are presented in Section \ref{sec:Space-filling designs}. 

\subsection{Standard Monte Carlo sampling}
\label{sec:Standard_sampling}

The natural method of constructing the sample set $\{ \bm{\xi}^{(i)} \}_{i=1}^{N}$, known as the standard MC sampling, is to generate $\bm{\xi}$ according to the orthogonality measure $f(\bm \xi)$ of the PC basis, for which the weights are $w(\bm{\xi}) = 1$. For instance, standard MC sampling of the $d$-dimensional Legendre polynomials corresponds to sampling from the uniform distribution on $[-1, 1]^d$. For other polynomials, the corresponding sampling distributions are listed in Table \ref{table:Askey}. 

As shown in \citep[Theorem 3.1]{Hampton15b}, a standard MC sampling of the $d$-dimensional Legendre polynomials of total order $p$ gives a coherence parameter $\mu \le \exp(2p)$, while $\mu$ associated with standard MC sampling of the $d$-dimensional Hermite polynomials is bounded by $\mu \le C_p \eta_p^p$, with $C_p$ and $\eta_p$ constants depending on $p$. Notice that these bounds are independent of the dimension $d$, thus suggesting the use of standard MC sampling when $d$ is large and the QoI admits a low order PCE. For high-order PCE's, $\mu$ is large and therefore $N$ needs to be large, according to Theorem \ref{thm:SampleDepth}. This is why, for those cases, alternative sampling distributions are needed to reduce $N$. Convergence analyses of the standard MC method for univariate and multivariate Legendre polynomials are also presented in \citep{Migliorati_2014} and \citep{Migliorati_2013b}, respectively.

\subsection{Coherence-optimal sampling}
\label{sec:Coherence-optimal sampling}

For a fixed sample size $N$, the coherence parameter $\mu$ plays a key role in the stability and convergence of least squares PC approximation, as indicated in (\ref{eqn:prob_stability_bound}) -- along with Corollary \ref{cor:stability} -- and Theorem \ref{thm:SampleDepth}. Heuristically, a smaller $\mu$ results in a more stable and accurate approximation. This has motivated the design of a random sampling strategy, dubbed {\it coherence-optimal} in \cite{Hampton15b}, which seeks to find a sampling measure -- instead of discrete sample points $\{ \bm{\xi}^{(i)} \}_{i=1}^{N}$ --  to minimize $\mu$. Specifically, given a fixed PC basis $\{\psi_j(\bm\xi)\}$, define 
\begin{equation} 
\label{eqn:B}
B(\bm \xi) := \sqrt{\sum_{j=1}^P |\psi_j(\bm \xi)|^2}.
\end{equation}
Notice that $B^2(\bm \xi)$ represent a uniformly least upper bound on the sum of the squares of the PC basis functions. As discussed in \cite{Hampton2015,Cohen16}, sampling inputs according to the alternative measure 
\begin{equation} 
\label{eqn:pdf_coh_opt}
f_{\bm Y}(\bm \xi) := c^2f(\bm \xi)B^2(\bm \xi)
\end{equation}
with the weight function $w(\bm\xi)=c^{-1}B(\bm\xi)^{-1}$ leads to the minimum possible $\mu=P$ compared to any other sampling measure. In (\ref{eqn:pdf_coh_opt}), $c^2=\left(\int_{\Omega}f(\bm\xi)B^2(\bm\xi)d\bm\xi\right)^{-1}=P^{-1}$ is a normalization constant. 
An analytic expression for $B(\bm\xi)$ is not generally available; however, for a given set of PC basis, $B(\bm\xi)$ can be evaluated at any arbitrary values of $\bm\xi$. This has motivated using a Markov Chain Monte Carlo approach to sample from $f_{\bm Y}(\bm\xi)$ in (\ref{eqn:pdf_coh_opt}). A detailed description of this technique can be found in \cite[Section 4.3.1]{Hampton15b} and \cite[Algorithm 1]{Hampton2015}. 

Following \citep{Hampton15b}, the coherence-optimal sampling approach ensures a stable computation of $\bm c$ with a number of solution realizations that depends linearly (up to a logarithmic factor) on the number of PC coefficients $P$, i.e., $N\sim\mathcal{O}(P\log P)$. Moreover, through the application of this approach to various numerical examples, it was empirically observed that the coherence-optimal sampling leads to either similar or considerably more accurate results in comparison to sampling from $f(\bm\xi)$ \citep{Hampton15b}. 

\subsection{Asymptotic sampling}
\label{sec:Asymptotic sampling}

For asymptotically large $p$, the coherence-optimal density $f_{\bm Y}(\bm\xi)$ associated with Legendre or Hermite polynomials, respectively, converges to or can be approximated with known distributions, \citep{Hampton2015,Hampton15b}. 
Specifically, for the case of Legendre polynomials, the asymptotic density coincides with that of i.i.d. Chebyshev random variables over $[-1,1]$, i.e., 
\begin{equation} 
\label{eqn:cheb_legendre}
f_{Y}(\xi_k) := \frac{1}{\pi \sqrt{1-\xi_k^2}},
\end{equation}
with the corresponding weights given by 
\begin{equation} 
\label{eqn:w_asym}
w(\bm \xi) := \prod_{k=1}^d (1-\xi_k^2)^{1/4}. 
\end{equation}
The coherence parameter of this asymptotic sampling is bounded by $\mu \le 3^d$, thus suggesting the advantage of Chebyshev sampling for high order PCE in low enough dimensions, \citep[Theorem 3.2]{Hampton15b}.

An asymptotic analysis of the Hermite polynomials which utilizes Hermite Functions suggests sampling uniformly from a $d$-dimensional ball of radius $\sqrt{2} \sqrt{2p+1}$, with weights given by $w(\bm \xi) := \exp(-\Vert \bm \xi \Vert^2_2/4)$, \citep[Section 3.2]{Hampton15b}. The corresponding coherence parameter is given by $\mu = \mathcal{O}({(2p)^{d/2}}/{\Gamma(d/2+1)})$, with $\Gamma$ being the Gamma function, which features a weaker dependence on $p$ in comparison to $d$. We refer the interested reader to \cite[Section 3.2]{Hampton15b} for a detailed description of the uniform sampling from a $d$-dimensional ball. 

\subsection{Subsampled quadrature points}
\label{sec:SC}

A widely used sampling-based approach to compute the PC coefficients is the stochastic collocation (SC) method \cite{Mathelin03, Xiu05a, Babuska05R, Constantine09,Constantine12a,Jones13}. The main idea behind the SC technique is to sample the output QoI at particular points, e.g., Gaussian quadrature points, in the stochastic space and then approximate the solution via interpolation or its (integral) statistics via quadrature integration. SC may also be used to compute the PC coefficients by approximating the $d$-dimensional  integrals (projection equations) in (\ref{eq:projection}) \cite{Constantine12a} using tensor product or sparse grids constructed from, for instance, one-dimensional Gaussian quadrature points \cite{Constantine12a}. This is known as pseudo-spectral approximation. 


Although it has been shown that SC based on full tensor product grids is an effective technique for low-dimensional stochastic problems \cite{Xiu05a,Babuska_2007}, it suffers from the so-called \textit{curse-of-dimensionality} for large input dimensions $d$: the number of collocation points $N$ in full tensor product grids grows exponentially fast as a function of $d$, i.e., $N=N_1^d$, where $N_1$ is the number of quadrature points in one dimension. For high-dimensional problems, sparse combination of tensor product rules, as for instance introduced by Smolyak \cite{Smolyak63}, have been proposed to alleviate the issue of curse-of-dimensionality \cite{Xiu05a,Nobile2008}. 

In the context of least squares PCE's, Zhou et al. \cite{Zhou_2015} proposed to generate $\{\bm{\xi}^{(i)}\}_{i=1}^{N}$ from $N$ randomly (uniformly) selected points of a tensor product Gaussian quadrature grid. It was shown in \cite{Zhou_2015} that this sampling approach, referred to as {\it randomized quadratures}, is asymptotically (in order) stable with a number of samples which scales linearly (up to a logarithmic factor) with $P$. Very recently, a technique based on deterministic selection of samples from full tensor grids is proposed in \cite{Seshadri16} as an alternative to the randomized quadrature approach. 

\subsection{Optimal design of experiments}
\label{sec:optimal ODE}

Planning an experimental procedure in order to determine the relation between a process inputs, a.k.a. factors, affecting the process outputs prior to performing the actual experiment is studied under the subject of {\it design of experiments} (DOE). Conducting experiments, either physical or computational, can be very expensive and time-consuming; hence, it is of interest to extract as much information as possible from a given amount of experimental effort, which is addressed in the context of ODE. One of the earliest work on ODE was presented in 1918 by Smith \cite{Smith_1918}, but the major contributions took place a few decades later when Kiefer and co-workers explained the theory behind ODE and proposed a framework for practical implementation of the $D$-optimal designs \cite{Brown_1985}. 

In this work we are interested in approximating the map between model inputs -- that are random -- and output QoIs in a basis of orthogonal polynomials. In particular, this is achieved via the LSA using selected realizations of inputs and outputs. Although in the context of least squares regression -- as opposed to least squares approximation (see Section \ref{sec:alphabetic_interp}) -- a selection of input realizations for building polynomial models with certain optimality goal has been expensively studied over the last few decades; see, e.g., \citep{Fedorov72, Fedorov1997, Box2005, Pukelsheim2006, Atkinson08}. The following section provides a brief review and interpretation of a major class of model-based ODE, known as alphabetic optimal design, in relation to least squares PC approximation. 

\subsubsection{Alphabetic optimality criteria}
\label{sec:Classical optimality criteria}

As discussed in Section \ref{sec:Stability}, the information matrix $\bm{M}$ plays a key role in the stability of the least squares solution $\hat{\bm c}$, described by its deviation from the identity, condition number, or largest singular value. This has motivated the development of several criteria in ODE to choose sample points $\{\bm{\xi}^{(i)}\}_{i=1}^{N}$ in a manner that some functional of $\bm M$, $\phi(\bm{M})$, known as alphabetic or classical optimality criterion, is minimized \citep{Morris10}. Among the most popular alphabetic optimality criteria, we mention:

\begin{itemize}
\item $D$-optimality: An optimal design obtained by minimizing the determinant of the inverse of information matrix, i.e., $\phi_{D}=|\bm{M}^{-1}|^{1/P}$.  

\item $A$-optimality: Takes the trace of the inverse of the information matrix as the overall measure of the average variance of the estimators. That is, $\phi_{A}=\text{Tr}( \bm{M}^{-1})$ is minimized, where $\text{Tr}(\cdot)$ denotes the trace operator. 

\item $E$-optimality: Minimizes the largest eigenvalue $\lambda_{\max}$ of the inverse of the information matrix, i.e., $\phi_{E}=\lambda_{\max}( \bm{M}^{-1})$ is minimized. 

\item $K$-optimality \cite{Chen11,Ye13}: Minimizes the condition number of the information matrix, i.e, $\phi_{K}=\mathcal{K}(\bm M)$.

\item $I$-optimality: Minimizes $\phi_I= \text{Tr} (\tilde{\bm{M}} \bm{M}^{-1})$, where $\tilde{\bm{M}}= \int_{\Omega} \bm{\psi}^{T}(\bm{\xi}) \bm{\psi}(\bm{\xi}) f(\bm{\xi}) d{\bm{\xi}}$ (see for example \cite[Page 5]{Hardin1993} or \cite[Page 473]{Myers_2009}) and $\bm{\psi}(\bm{\xi})$ is a row vector containing the PC basis functions. Due to the orthonormality of the PC basis, $\tilde{\bm M}$ is the $P\times P$ identity matrix and, therefore, $I$-optimality is the same as  $A$-optimality.
\end{itemize}

\begin{remark1}
Instead of minimizing $\phi_{D}$, $\phi_{A}$ and $\phi_{E}$, one may alternatively maximize $|\bm{M}|^{1/P}$, $\text{Tr}( \bm{M})$, $\lambda_{\min}( \bm{M})$ to obtain $D$-, $A$- and $E$-optimal designs, respectively \citep{AndersonCook2009, Fedorov1997}. 
\end{remark1}

In the ODE literature, \textit{D}-, \textit{A}-, and \textit{E}-optimal designs are referred to as estimation-oriented optimal designs \citep{GOOS2012}, meaning that these designs are focused on precise estimation of the coefficients ${\bm{c}}$. On the other hand, prediction-oriented optimal designs, such as $I$-optimal, a.k.a. \textit{Q}-optimal \cite{Fedorov72}, or \textit{V}-optimal \cite{AndersonCook2009}, aim at reducing the variance of the approximation $\hat{u}$. $I$-optimal designs have received increasing attention in the context of response surface designs, where the ultimate goal is to make predictions \citep{GOOS2012}. There are more optimality criteria that are labeled by other letters in the literature and have been introduced over the time. A discussion on those optimal measures is beyond the scope of the present work and we refer the interested reader to  \cite{ Kiefer_1959,Fedorov1997,Pukelsheim2006,Atkinson08, GOOS2012} for more details.  

\subsubsection{Interpretation of alphabetic optimality criteria in statistical regression setting}
\label{sec:alphabetic_interp}

We highlight a distinction between the least squares approximation of this work and those in the statistical regression literature, despite both using the same least squares construction as in (\ref{eqn:regression2}). In the latter setting, the vector $\bm\epsilon$ models the uncertainty or noise in the data/measurements and is often modeled by a random vector with known statistics, e.g., i.i.d. zero mean Gaussian vector, and $\bm\xi$ is assumed to be deterministic. In the former, however, $\bm \xi$ represents the uncertainty in the underlying model and, given the samples $\{\bm{\xi}^{(i)}\}_{i=1}^{N}$, $\bm \epsilon$ is a deterministic error that is unknown. The interpretation of the model (\ref{eqn:regression5}) in the statistical regression setting leads to particular interpretations of the alphabetic optimality criteria. For example, $D$- and $A$-optimal designs minimize the geometric and arithmetic mean of the variation in the estimated coefficients $\hat{\bm c}$, respectively, \citep{Faller03}. $I$-optimal design minimizes the average variance of $\hat{u}$ over possible realizations of $\bm\xi$. These variations are a consequence of the uncertainty in the data introduced by $\bm\epsilon$.

\subsubsection{Constructing optimal designs}
\label{sec:Constructing optimal designs}

In Section \ref{sec:introduction}, we categorized ODE under the deterministic sampling techniques. This is in fact true only for limited cases of low-order polynomial models with a small number of inputs for which analytic expressions for the optimal designs are available, see, e.g., \cite{Yang_2012, dette_2014,Kabera_2015}. In general, however, the ODE optimization problems are solved numerically, often, involving search through a large, but finite, number of randomly generated candidate designs. The selected optimal design, therefore, depends on the choice of the candidate ensembles. This suggests that one may consider ODE as (quasi-)random sampling.

Heuristic exchange algorithms are among the earliest search methods proposed to construct optimal designs \cite{Smucker_2010,Mandal_2015}. Exchange algorithms were originally developed for $D$-optimal designs since, due to the availability of an update equation for the determinant of the information matrix (see \ref{apx:det_update}), they are computationally more feasible in comparison to other criteria \cite{Fedorov1997}. Moreover, it has been shown that $D$-optimal designs perform well relative to other criteria \cite{Atkinson08}.

The original Fedorov exchange algorithm (FEA) proposed in \cite{Fedorov72} starts by generating a discrete candidate set which includes $N_c > N$ realizations of  $\bm\xi$, generated for instance according to $f(\bm\xi)$, and the corresponding $N_c \times P$ candidate measurement matrix $\bm \Psi_c$. Next, a non-singular initial design $\bm \Psi$ of size $N \geq P$ is selected randomly from $\bm \Psi_c$ and FEA iteratively modifies the current design by exchanging a row $\bm \psi_i$ in $\bm \Psi$ with a row $\bm \psi_j$ in $\bm \Psi_c$ which corresponds to the maximum reduction in the $D$-optimal measure $\phi_{D}=|\bm{M}^{-1}|^{1/P}$ of $\bm\Psi$. Let $\bm \Psi_{\text{old}}$ and $\bm \Psi_{\text{new}}$ denote the measurement matrix $\bm \Psi$ before and after the exchange, respectively. Using (\ref{eqn:update_det}) the update formula for the determinant is
\begin{equation}
\label{eqn:update_fed}
\frac{|\bm \Psi^T_{\text{new}} \bm \Psi_{\text{new}}|}{|\bm \Psi^T_{\text{old}} \bm \Psi_{\text{old}}|}=1 + \Delta(\bm \psi_i,\bm \psi_j),
\end{equation}
where the Fedorov's $\Delta$-function is given by 
\begin{eqnarray}
\label{eqn:delta_fed}
&& \Delta(\bm \psi_i,\bm \psi_j)=d(\bm \psi_j) - \left[ d(\bm \psi_i) d(\bm \psi_j) - (d(\bm \psi_i, \bm \psi_j))^2\right] -d(\bm \psi_i); \\
&& d(\bm \psi_i) = \bm \psi^T_i(\bm \Psi^T_{\text{old}} \bm \Psi_{\text{old}})^{-1} \bm \psi_i; \nonumber  \\
&&  d(\bm \psi_i, \bm \psi_j)=\bm \psi^T_i(\bm \Psi^T_{\text{old}} \bm \Psi_{\text{old}})^{-1} \bm \psi_j. \nonumber
\end{eqnarray}
Consequently, when the $D$-optimal design is constructed by maximizing $|\bm{M}|^{1/P}$, the couple $(\bm \psi_i,\bm \psi_j)$ which correspond to the largest value of the Fedorov's $\Delta$-function are exchanged to provide the largest improvement in the optimality criterion. The algorithm stops when the largest value of the $\Delta$-function is smaller than a user specified threshold, e.g., $10^{-6}$ \cite{Fedorov72}. Convergence proof of the FEA can be found in \cite{Fedorov72}. 

At each iteration of the original FEA only one exchange is performed. A modified version of FEA with multiple exchanges in each iteration was proposed in \cite{Cook_1980}. Other popular exchange algorithms include DETMAX algorithm \cite{Mitchell_1974}, Wynn's algorithm \cite{Wynn_1970}, $k$-exchange algorithm \cite{Johnson_1983}, $kl$-exchange algorithm \cite{Atkinson_1989}, and coordinate exchange algorithm of \cite{Meyer_1995} which does not require a candidate set. Performance comparison between different exchange algorithms can be found in \cite{Cook_1980,Nguyen_1992, Johnson_1983, Pronzato_2008}. A discussion on details of these methods is beyond the scope of the present work and we refer the interested reader to the above references for further details. 

For either high-dimensional problems or when high order PCE's are required, the number of unknown coefficients $P$ may be very large. Consequently, the candidate measurement matrix should have a large number of rows to avoid sub-optimal solutions. This results in a large number of exchanges to be performed and eventually high computational costs. Moreover, for criteria other than $D$-optimality, there is no update formula similar to (\ref{eqn:update_fed}) for the simultaneous inclusion and exclusion of the design points that can be used to expedite the exchange procedure. 

In order to relax the computational complexity of constructing the alphabetic optimal designs, sequential (greedy) augmentation techniques have been developed \citep{Dykstra_1971,Song_2009,Shin_2016,Burnaev2016}. In particular, starting from a (non-singular) design with $n< N$ rows, Dykstra proposed to choose the $(n+1)-th$ row from the candidate design $\bm \Psi_c$ such that the $D$-optimality criterion is maximally improved \citep{Dykstra_1971}. As no exchanges are performed, this algorithm is considerably faster than FEA. Moreover, since only rows are being added to the design, one may employ the simple update formula for the trace of $\bm M^{-1}$ (see \ref{apx:trace_update}) to accelerate the construction of $A$-optimal designs via sequential augmentation. In fact, this technique has been employed in \cite{Song_2009} to construct $A$-optimal Bayesian designs. There, it was suggested that the sequentially generated optimal designs may be improved further by FEA, but a noticeable improvement was not observed for the problem at hand \cite{Song_2009}.  A similar sequential algorithm has been used in \cite{Shin_2016} to construct designs that maximize the optimality criterion $\mathcal{S}(\bm{\Psi}) := \left(\sqrt{|\bm{\Psi}^T\bm{\Psi}|}/ \prod_{j=1}^P \Vert \bm{\Psi}(:,j) \Vert_2 \right)^{1/P}$, where $\bm{\Psi}(:,j)$ is the $j$th column of $\bm{\Psi}$. In that work, when $n< P$ rows are selected, only the first $n$ columns of $\bm\Psi$ are considered; otherwise, $\bm\Psi$ (hence $\bm M$) will be singular. Stated differently the rows and columns of $P$ are increased simultaneously by one until $P$ rows are selected. When $n\ge P$, only new rows are added until $N$ rows are selected. Very recently, an accurate estimation of the PCE-based Sobol' indices via the LSA technique has been reported in \cite{Burnaev2016}, where the $D$-optimal sample set is generated using a similar greedy approach. 

In the present work, we employ the sequential augmentation scheme to construct $A$-, $E$-, and $D$-optimal designs as outlined in Algorithm \ref{alg:sequential algorithm}. The following points are worth highlighting regarding this algorithm: 

\begin{itemize}
  \item As new rows are added to the design, the computation of $D$- and $A$-optimality criteria may be accelerated by employing the update formulas given in (\ref{eqn:update_det}) and (\ref{eqn:update_trace}), respectively. There is no update formula for the $E$-optimal designs, hence, its construction is considerably more expensive. 
  \item Repeating samples are not allowed as they lead to singular designs. Hence, it is crucial to search among the remaining rows of $\bm \Psi_c$ when adding new rows to $\bm\Psi$. 
\item In general, the exact solution to alphabetic optimal design is NP-hard \cite{Mandal_2015} and thus infeasible for large $P$. A design obtained by the greedy Algorithm \ref{alg:sequential algorithm} may therefore be sub-optimal. Moreover, the alphabetic optimality criteria $\phi$ are non-convex and typically exhibit multiple local minima/maxima. In order to increase the chance of finding the global optima, it is advised to run them multiple times with different (random) initial designs. 
\end{itemize}

\begin{algorithm}[H]
\caption{{\small A sequential algorithm to construct alphabetic optimal designs. Adopted from \cite{Dykstra_1971, Shin_2016, Song_2009}.}}
\label{alg:sequential algorithm}
{\small
\begin{algorithmic}[1]
\STATE $\bullet$ \textbf{Input}: Number of samples in the design $N\ge P$, number of samples in the candidate design $N_c$, optimality criterion $\phi$, type, dimension $d$, and order $p$ of PC basis.
\STATE $\bullet$ \textbf{Output}: $N \times P$ optimal measurement matrix $\bm \Psi$ and the corresponding optimal sample set $\{ \bm{\xi}^{(i)} \}_{i=1}^{N}$ based on the smallest optimality criterion $\phi$.
\STATE Generate $N_c$ realizations $\{ \bm{\xi}^{(i)} \}_{i=1}^{N_c}$ and the corresponding $N_c \times P$ candidate measurement matrix $\bm \Psi_c$.
\STATE Initialize $\mathcal{I}=\{\emptyset\}$. Let $\bm \Psi_c(\mathcal{I},\mathcal{J})$ be the sub-matrix of $\bm \Psi_c$ with row and column indices in $\mathcal{I}$ and $\mathcal{J}$, respectively.
\FOR{$n=1:N$}
\STATE Grow the column set while $n< P$ by setting $\mathcal{J}=\{1,\cdots,\min(n,P)\}$.
\FOR{$i\in\{1,\cdots,N_c\}\backslash\mathcal{I}$}
\STATE $\phi_i = \phi\left(\bm\Psi_c(\mathcal{I}\cup\{i\},\mathcal{J})\right)$.
\ENDFOR
\STATE Include the row corresponding to largest reduction in $\phi$, i.e., set $\mathcal{I}=\mathcal{I}\cup \{\argmin_{i} \phi_i\}$. 
\ENDFOR
\STATE Return $\bm\Psi = \bm\Psi_c(\mathcal{I},\mathcal{J})$
\end{algorithmic}
}
\end{algorithm}

In addition to the greedy algorithms discussed here, other techniques such as simulated annealing (SA) \cite{Haines_1987, Meyer_1988}, genetic algorithms (GA) \cite{Montepiedra_1998, Broudiscou_1996}, semi-definite programming (SDP) \cite{Sagnol_20111, Atashgah_2008} and pattern search \cite{Hooke_1961,Hardin1993} have been employed in the literature to construct alphabetic optimal designs. Moreover, software packages such as JMP \cite{Goos11}, POBE \cite{Maus_2014}, AlgDesign \cite{Wheeler_2004}, PopED \cite{Foracchia_2004,Nyberg_2012} and Gosset \cite{gosset} have been developed to construct ODE using the algorithms discussed here. We note that the majority of the software packages for the construction of ODE are developed for scenarios where low-order (mostly quadratic) polynomials are employed, hence, limiting the applicability of such packages for high-order PCE's.  

\subsection{Space-filling designs}
\label{sec:Space-filling designs}

Space-filling designs form a class of DOEs that have received extensive attention in the context of computer experiments \cite{Santner_2003,Kleijnen_2008, Pronzato_2011, Joseph_2016}. The main idea behind space-filling designs, as suggested by the name, is to evenly spread input samples over the sample space $\Omega$. This is motivated by the heuristic that when the realization of a QoI at an input parameter is deterministic (this may not hold for physical experiments), evaluating the QoI at closely-spaced points may not lead to new information about the QoI, particularly in the scenarios where only a small number of realizations can be afforded.  In the following, we will review two important space-filling designs, namely quasi-Monte Carlo (QMC) and Latin hypercube (LH) sampling techniques. 

\subsubsection{Quasi-Monte Carlo sampling}
\label{sec:QMC_sampling}

To alleviate the slow convergence rate of standard MC technique, especially for high-dimensional integrations in finance applications \citep{Wang_2008}, researchers have proposed a deterministic version of the MC sampling method based on the {\it low-discrepancy} sequences \citep{Dick_2010} known as {\it quasi-Monte Carlo} (QMC) technique \citep{niederreiter_1978,caflisch_1998, Blatman_2007,Migliorati15}. It has been shown that for integrating functions of $d$-dimensional uniform inputs, a convergence rate of approximately $\mathcal{O}(N^{-1}(\log N)^d)$, which is asymptotically better than that of MC, can be expected from the QMC technique \citep{caflisch_1998}. 

Unlike {\it pseudo-random} sequences used in standard MC, low-discrepancy sequences, a.k.a. {\it quasi-random}, such as Hammersley set, Niederreiter, Halton, Faure, and Sobol sequences \citep{McLeish_2011}, are constructed such that the maximal degree of uniformity is provided for the sample points drawn from a uniform distribution. A quantitative assessment of the degree of uniformity of a set of sample points is achieved by calculating its {\it discrepancy}. Among the many definitions proposed for the discrepancy of a sequence, star discrepancy $D_N^*$, defined as follows, is one of the most widely used ones due to its simplicity.

Star discrepancy is used in the context of low-discrepancy sequences for quasi-Monte Carlo sampling techniques. Given the sample set $\{ \bm{\xi}^{(i)} \}_{i=1}^{N}$ in $d$-dimensional unit cube $[0,1)^d$, the star discrepancy $D_N^*$ of the set $\{ \bm{\xi}^{(i)} \}_{i=1}^{N}$ is defined as 
\begin{equation} 
\label{eqn:star-disc}
D_N^* := \sup_{J \in \mathcal{J}^*} \left\vert \frac{1}{N} \cdot \# \{i: \bm{\xi}^{(i)} \in J, 1 \le  i \le  N \} - \lambda_d(J) \right\vert,
\end{equation}
where $\mathcal{J}^*$ is the class of all subintervals $J$ of $[0,1)^d$ given by 
\begin{equation} 
\label{eqn:star-disc}
J = \prod_{k=1}^d [0,v_k)^d, \qquad 0 \le  v_k \le 1,  
\end{equation}
$\# \mathcal{A}$ denotes the number of elements of the set $\mathcal{A}$, and $\lambda_d(J)$ is the Lebesgue measure of $J$ \citep{Entacher_2001}. A well uniformly distributed set of points has a small star discrepancy \citep{niederreiter_1978,caflisch_1998}.

In \citep{Migliorati15}, a stable and accurate solution to the LSA problem in (\ref{eqn:regression1}) is achieved with low-discrepancy sample points and a number of samples $N$ which is proportional to the square of $P$ (up to a logarithmic factor). To the best of our knowledge, the application of QMC sampling technique for least squares PCE's has not been extended to problems with non-uniform random inputs. For a detailed description of QMC sampling approach for multivariate Legendre polynomials we refer the interested reader to \citep{Migliorati15}.

Another class of space-filling designs, known as {\it uniform} design, which facilitate the same goal of constructing samples with maximum degree of uniformity was proposed in \cite{Fang_1980}. Uniform designs are constructed based on alternative and computationally less expensive measures of the uniformity rather than the star discrepancy $D_N^*$ used in QMC \cite{Fang_2000}. Several methods such as good lattice \cite{Fang_1994}, optimization searching \cite{Winker_1998}, collapsing  \cite{Fang_2003}, combinatorial construction \cite{zhang_2005}, and Latin square \cite{Fang_1999} have been proposed to construct uniform designs.

\subsubsection{Latin hypercube sampling}
\label{sec:LH sampling}

Latin hypercube (LH) sampling, first introduced in \cite{McKay_1979}, aims at selecting the sample set $\{ \bm{\xi}^{(i)} \}_{i=1}^{N}$ such that these points cover all portions of the design space $\Omega$. LH sampling attempts to reduce the variance of standard MC estimators by selecting $N$ sample points from $N$ equiprobable partitions (hypercubes) of the sample space, which is known as {\it stratification} of the probability distribution. In \cite{Stein_1987, Owen_1998}, it is shown that the convergence rate of the LH sampling is never worse than that of standard MC when $N \rightarrow \infty$. However, under some conditions discussed in \cite{McKay_1979}, e.g., when certain monotonicity conditions do not hold, LH may not be advantageous over the standard MC in reducing the variance of an estimator. Algorithm \ref{alg:LHS} summarizes the main steps to generate $\{ \bm{\xi}^{(i)} \}_{i=1}^{N}$ via the LH sampling technique. 

\begin{algorithm}[H]
\caption{{\small Summary of the Latin hypercube sampling approach.}} 
\label{alg:LHS}
\begin{algorithmic}[1]
{\small
\STATE Divide the pdf of each random input $\Xi_{k}$  into $N$ equiprobable intervals.
\STATE Sample the $i$th interval uniformly by $z_i = \frac{i-1}{N} + \frac{\zeta_i}{N}$, where $\zeta_i$ is a uniform random number over $[0,1]$ drawn independently. Repeat this (independently) for all intervals (totally $N$).
\STATE Compute the $i$th realization of random variable ${\Xi}_{k}$, $\xi_k^{(i)}$, via the inverse of its cumulative
distribution function (CDF), i.e., $\xi_k^{(i)} = F_{{\Xi}_{k}}^{-1}(z_i)$. Repeat this for all $z_i$.
\STATE Repeat 2-3 for all the inputs $\Xi_k$.
\STATE Randomly pair the $N$ realizations of the inputs so that $\bm{\xi}^{(i)}=(\xi_1^{(i_1)},\cdots,\xi_d^{(i_d)})$, where, for each $k$, $i_k$ is selected randomly (without replacement) with equal probability from the set $\{1,\cdots,N\}$, i.e., $\{i_k\}$ is a random permutation of $\{1,\cdots,N\}$.
}
\end{algorithmic}
\end{algorithm}

Several modified versions of LH sampling such as orthogonal \cite{Ye_1998,  Ye_2000,Butler_2001} and column-wise orthogonal  \cite{Steinberg_2006}, sliced and optimal sliced \cite{Qian_2012}, and orthogonal-maximin \cite{Joseph_2008} have been proposed to improve the efficiency and accuracy of the LH sampling approach. Algorithm \ref{alg:LHS} is used when the random inputs are independent from each other. For the correlated random inputs, some extended/modified versions of LH sampling have also been proposed in the literature \cite{Iman_1982,Huntington_1998, Sallaberry_2008}.  

In addition to the space-filling designs we reviewed here, many other have been introduced over the past few decades. Among them we mention maximin designs \cite{JOHNSON_1990}, a.k.a. sphere packing designs, in which the minimum distance between pairs of designs points is maximized, maximum entropy designs \cite{Shewry_1987}, which takes the entropy as a measure of the uniformity of the samples, Gaussian process integrated mean squared-error (IMSE) designs minimizing the variance of prediction \cite{Sacks_1989, Sacks_1989b}, and maximum projection designs \cite{Joseph_2015}. 

\begin{remark1}
Although alphabetic optimal designs are developed for real world physical experiments, but these designs have also been successfully employed for computer experiments. As an example, we mention the work in \cite{Johnson_2010}, which compares $D$- and $I$-optimal designs with space-filling designs such as Latin hypercube, uniform, sphere packing, and maximum entropy, in terms of reducing the prediction variance for high-order polynomials. It was shown that $D$- and $I$-optimal designs result in prediction variances smaller than those obtained by space-filling designs. A similar comparison was also made in \cite{Johnson_2008}.  
\end{remark1}

\section{Numerical examples}
\label{sec:Numerical examples}
In this section, we study the empirical performance of six different sampling methods by considering three numerical examples. In the first example, the recovery of manufactured PCE's is investigated. Next, a nonlinear Duffing oscillator under free vibration which requires high order PC expansions is considered. Finally, we examine the application of a subset of the reviewed sampling techniques to the prediction of the remaining useful life of a Lithium-ion battery. 

We consider standard MC (Section \ref{sec:Standard_sampling}), coherence-optimal (Section \ref{sec:Coherence-optimal sampling}) and Latin hypercube (Section \ref{sec:LH sampling}) techniques denoted by {\it Standard}, {\it coh-opt} and {\it LH} sampling, respectively. In addition, we introduce hybrid sampling methods, denoted by {\it alphabetic-coherence-optimal}, that combine coherence-optimal technique with alphabetic optimality criteria. To construct these sample sets, the $N_c$ realizations of $\bm \Xi$ in the third line of Algorithm \ref{alg:sequential algorithm} are generated via the coherence-optimal technique of Section \ref{sec:Coherence-optimal sampling}. Consequently, the corresponding $N_c \times P$ candidate measurement matrix $\bm \Psi_c$ to be used in the sequential construction of $D$-, $A$-, and $E$-optimal designs will be already optimal in a sense that it yields to a small coherence parameter $\mu$, hence the names {\it D-coh-opt}, {\it A-coh-opt}, and {\it E-coh-opt}, respectively. We note that the larger $N_c$ in Algorithm \ref{alg:sequential algorithm}, the higher the quality of the constructed design and the associated computational cost. 

To verify the accuracy of the least squares PC model constructed with $N$ independent samples, we compute the relative error of the resulting PC expansion in predicting independent realizations of the QoI as
\begin{equation}
\label{eqn:rel_error}
\text{relative error}= \frac{\Vert \bm{u}_v - \bm{\Psi}_v \hat{\bm{c}} \Vert_2}{\Vert \bm{u}_v \Vert_2},
\end{equation}
where $\bm{u}_v$ is the vector of $N_v$ additional realizations of QoI (not used in computing $\hat{\bm c}$) and $\bm{\Psi}_v$ is the measurement matrix corresponding to $\bm{u}_v$. 

\begin{remark1}
\label{remark_average}
Because of the random nature of the discussed sampling techniques, the results presented in the following are obtained by averaging 60 independent replications using $N$ samples to compute $\bm c$ and $N_v$ realizations to evaluate the relative error in (\ref{eqn:rel_error}).
\end{remark1}
\subsection{Manufactured functions}
\label{sec:MF}

In the first example, we consider the reconstruction of manufactured PC expansions, where the coefficients $\bm c$ are prescribed to generate $\bm u$. Specifically, we consider the model 
\begin{equation} 
\label{eqn:PC_expansion_MF}
{u}(\bm{\xi}^{(i)}) = \sum_{j=1}^{P} c_j \psi_{j}(\bm{\xi}^{(i)}), \qquad i=1, \cdots, N,
\end{equation}
where each $c_j$ is generated independently from a standard normal distribution. Each realization $u(\bm\xi^{(i)})$ is then generated by evaluating the right-hand-side of (\ref{eqn:PC_expansion_MF}) at a set of $N$ samples of $\bm \xi$. Additionally, we assume that there is independent, normally distributed additive noise in the evaluation of ${u}(\bm{\xi}^{(i)})$ with a standard deviation of $0.03 \cdot |{u}(\bm{\xi}^{(i)})|$.  The PC expansion is then recovered successfully if the least squares solution yields a relative error $\le 0.02$.

\begin{figure}[ht]
\begin{center}
{\includegraphics[width = 16.5cm]{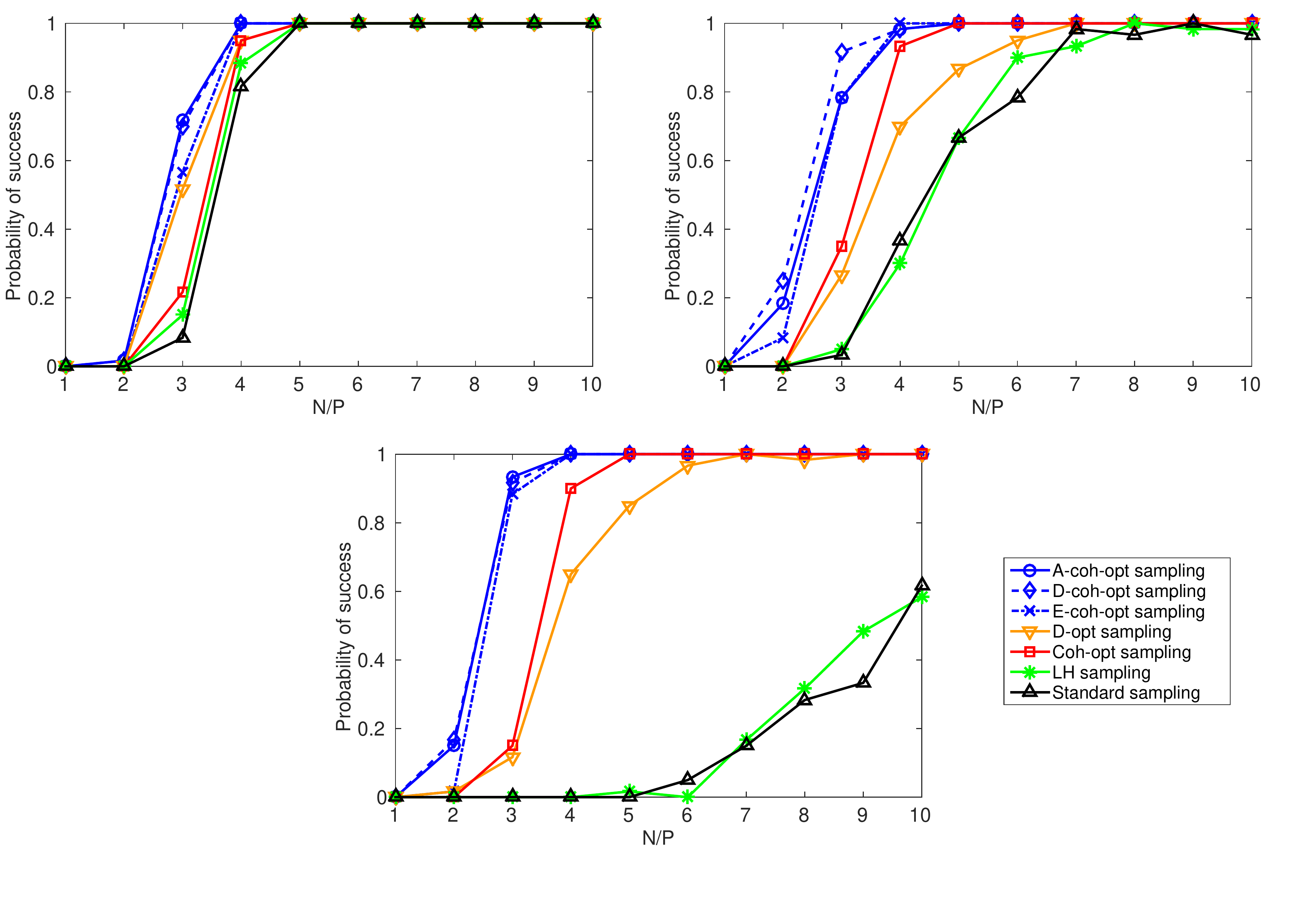}}
\put(-358,170){\text{\footnotesize (a)}}
\put(-123,170){\text{\footnotesize (b)}}
\put(-241,16){\text{\footnotesize (c)}}
\caption{Legendre recovery probability as a function of sample size $N$. (a) $(p,d)=(2,15)$; (b) $(p,d)=(4,4)$; (c) $(p,d)=(15,2)$.}
\label{fig:MF_Legendre}
\end{center}
\end{figure}

\begin{figure}[ht]
\begin{center}
{\includegraphics[width = 16.5cm]{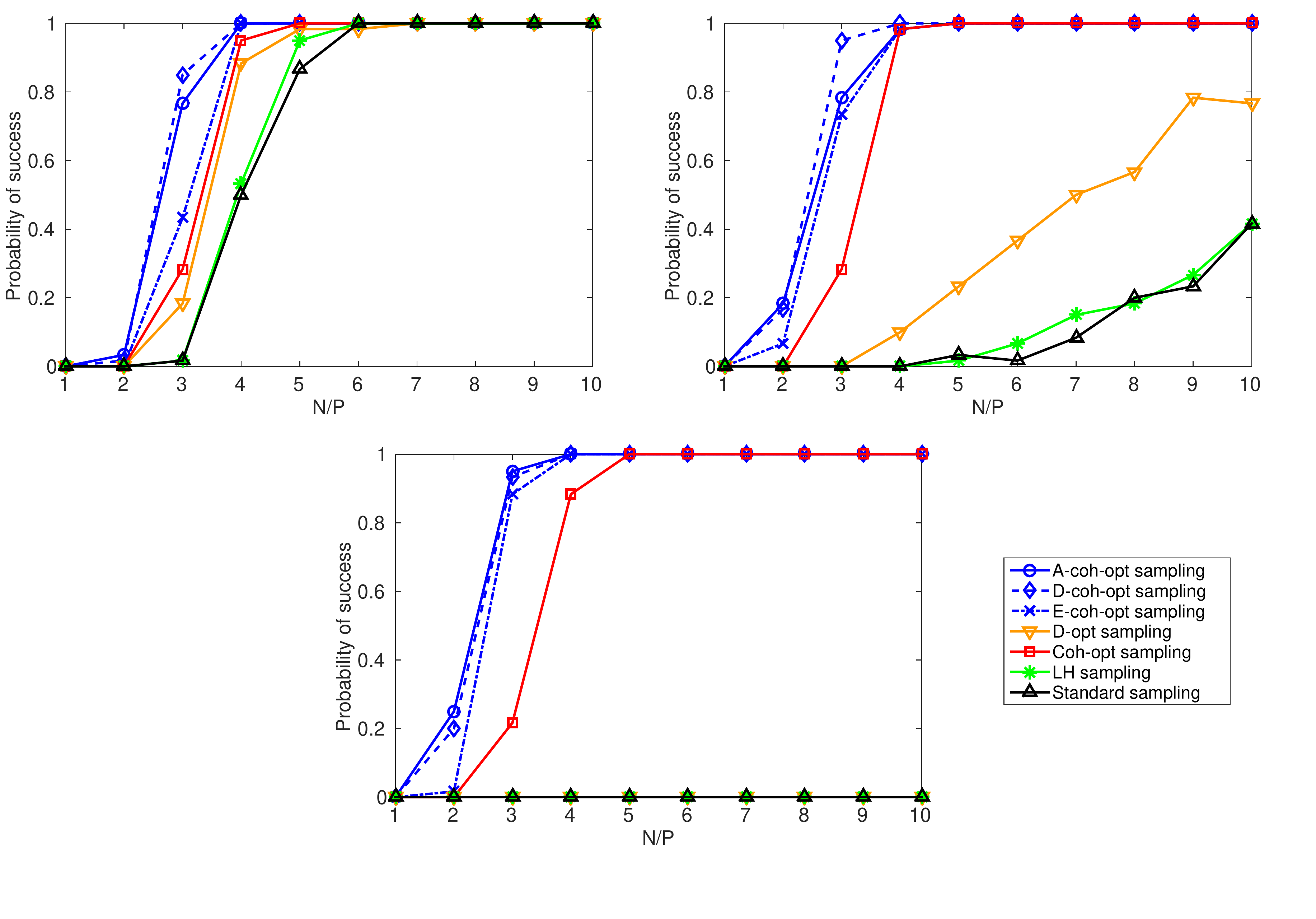}}
\put(-358,170){\text{\footnotesize (a)}}
\put(-123,170){\text{\footnotesize (b)}}
\put(-241,16){\text{\footnotesize (c)}}
\caption{Hermite recovery probability as a function of sample size $N$. (a) $(p,d)=(2,15)$; (b) $(p,d)=(4,4)$; (c) $(p,d)=(15,2)$.}
\label{fig:MF_Hermite}
\end{center}
\end{figure}

Figs. \ref{fig:MF_Legendre} and \ref{fig:MF_Hermite} represent the probability of a successful recovery of the manufactured solution as a function of the number of samples $N$ to the number of unknowns $P$ for the case of Legendre and Hermite polynomials, receptively. Three different scenarios, i.e., a high-dimensional case with $(p,d)=(2,15)$, a moderate case of $(p,d)=(4,4)$, and a high-order case with $(p,d)=(15,2)$ are considered in these figures. In this example, we consider $N_c = 4N$ for each value of the oversampling ratio $N/P$. The following observations are worth highlighting regarding Figs. \ref{fig:MF_Legendre} and \ref{fig:MF_Hermite}: 

\begin{itemize}
    \item For the high-order case $(p,d)=(15,2)$, both standard and LH sampling techniques demonstrate a poor performance in recovering the solution such that for the case of Hermite polynomials, they both fail to recover the solution even for large sample sizes, e.g., $N=10P$. These two methods perform significantly better for the high-dimensional, low-order case $(p,d)=(2,15)$, but are yet worse than the other sampling techniques for the moderate case $(p,d)=(4,4)$. 
    \item Coherence-optimal approach results in a more accurate solution recovery in comparison to the standard and LH sampling methods, especially, for the high-order case $(p,d)=(15,2)$. 
    \item In all cases, alphabetic-coherence-optimal techniques outperform the other sampling methods. They are significantly better than the standard and LH techniques for the high-order case $(p,d)=(15,2)$. Their advantage becomes more significant when the oversampling ratio $N/P$ is small. 
    \item Among the alphabetic-coherence-optimal techniques, {\it D-coh-opt} is slightly better than {\it A-coh-opt}, while {\it E-coh-opt} is the worst of three. 
     \item The main difference between the alphabetic-coherence-optimal technique and the alphabetic optimal design is that in the former approach the candidate ensemble $\bm \Psi_c$ is generated based on the coherence-optimal technique while in the latter $\bm \Psi_c$ is constructed via random sampling from the orthogonality measure $f(\bm \xi)$. To highlight the importance of constructing $\bm \Psi_c$ via coherence-optimal approach, we also report the results of the classical $D$-optimal design, denoted by {\it D-opt} in Figs. \ref{fig:MF_Legendre} and \ref{fig:MF_Hermite}, in which the candidate measurement matrix of Algorithm \ref{alg:sequential algorithm} is constructed by drawing samples from $f(\bm \xi)$. Although {\it D-opt} sampling performs fairly good for the the high-dimensional case in comparison to the alphabetic-coherence-optimal techniques, its accuracy is drastically reduced for the higher order expansions considered. 
\end{itemize}

\subsection{Nonlinear Duffing oscillator}
\label{sec:Duffing}

The second problem of interest in this study is to quantify the uncertainty in the displacement solution $u(\bm \Xi, t)$ of a nonlinear single-degree-of-freedom Duffing oscillator \cite{Mai_2015} under free vibration described by,
\begin{eqnarray} 
\begin{aligned}
& \ddot{u}(\bm \Xi, t) + 2 \omega_1 \omega_2 \dot{u}(\bm \Xi, t) + \omega_1^2 ( u(\bm \Xi, t) + \omega_3 u^3(\bm \Xi, t) )=0, \\
& {u}(\bm \Xi, 0) = 1, \qquad  \dot{u}(\bm \Xi, 0) = 0,
\end{aligned}
\label{eqn:duffing}
\end{eqnarray}
where the uncertain parameters $\{ \omega_i \}_{i=1}^3$, influencing the motion of the oscillator, are given by 
\begin{eqnarray} 
\label{eqn:duffing_rvs}
\begin{aligned}
& \omega_1 = 2 \pi \ (1 + 0.2 \ \Xi_1), \\ 
& \omega_2 = 0.05 \ (1+0.05 \ \Xi_2), \\
& \omega_3 = -0.5 \ (1+0.5 \ \Xi_3), 
\end{aligned}
\end{eqnarray}
with $\{ \Xi_i \}_{i=1}^3$ being i.i.d. uniform random variables $U(-1,1)$. 

\begin{figure}[H]
\begin{center}
{\includegraphics[width = 8.25cm]{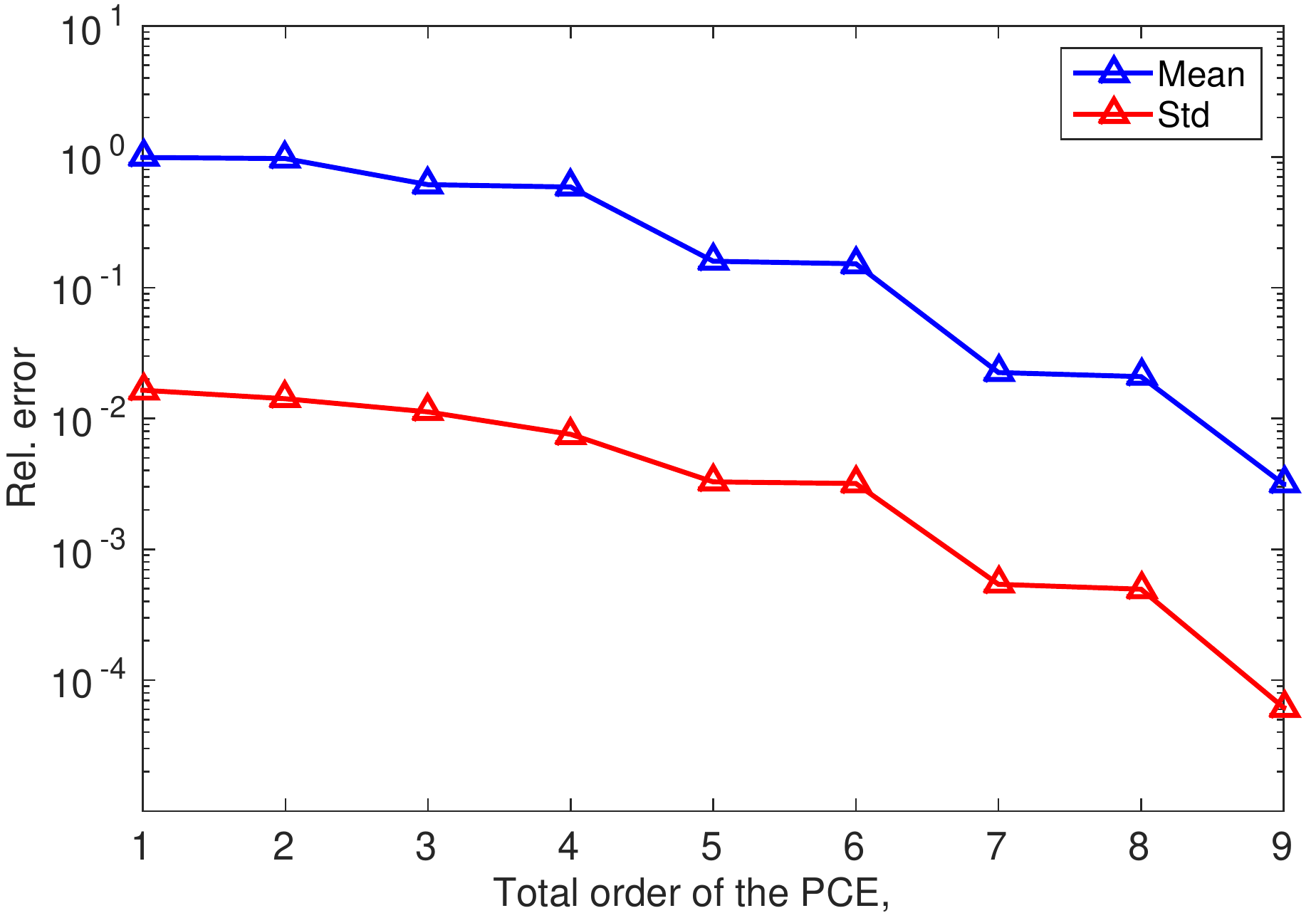}}
\put(-74.5,1.76){{\footnotesize $p$}}
\caption{Mean and standard deviation of the relative error of PCE of $u(\bm \Xi, 4)$ obtained via $N=20P$ standard MC samples.}
\label{fig:duffing_rand_gen}
\end{center}
\end{figure}

While this is a relatively low-dimensional problem ($d=3$), due to the presence of nonlinearities, high order PC expansions are required in order to maintain the approximation's accuracy at large time instances $t$. This high-order approximation requirement is illustrated in Fig. \ref{fig:duffing_rand_gen}, where we report the mean and standard deviation of the relative error of PCE of $u(\bm\Xi,4)$ obtained using the standard MC samples. In this experiment, we assumed $N=20P$ to generate sample size independent solutions. We considered an additional $N_v = 10000$ validation samples to compute the relative error in (\ref{eqn:rel_error}).

We considered $p=9$ for this problem to examine the performance of various sampling strategies. The pair $(p,d)=(9,3)$ leads to $P=220$ PC coefficients to be approximated. 

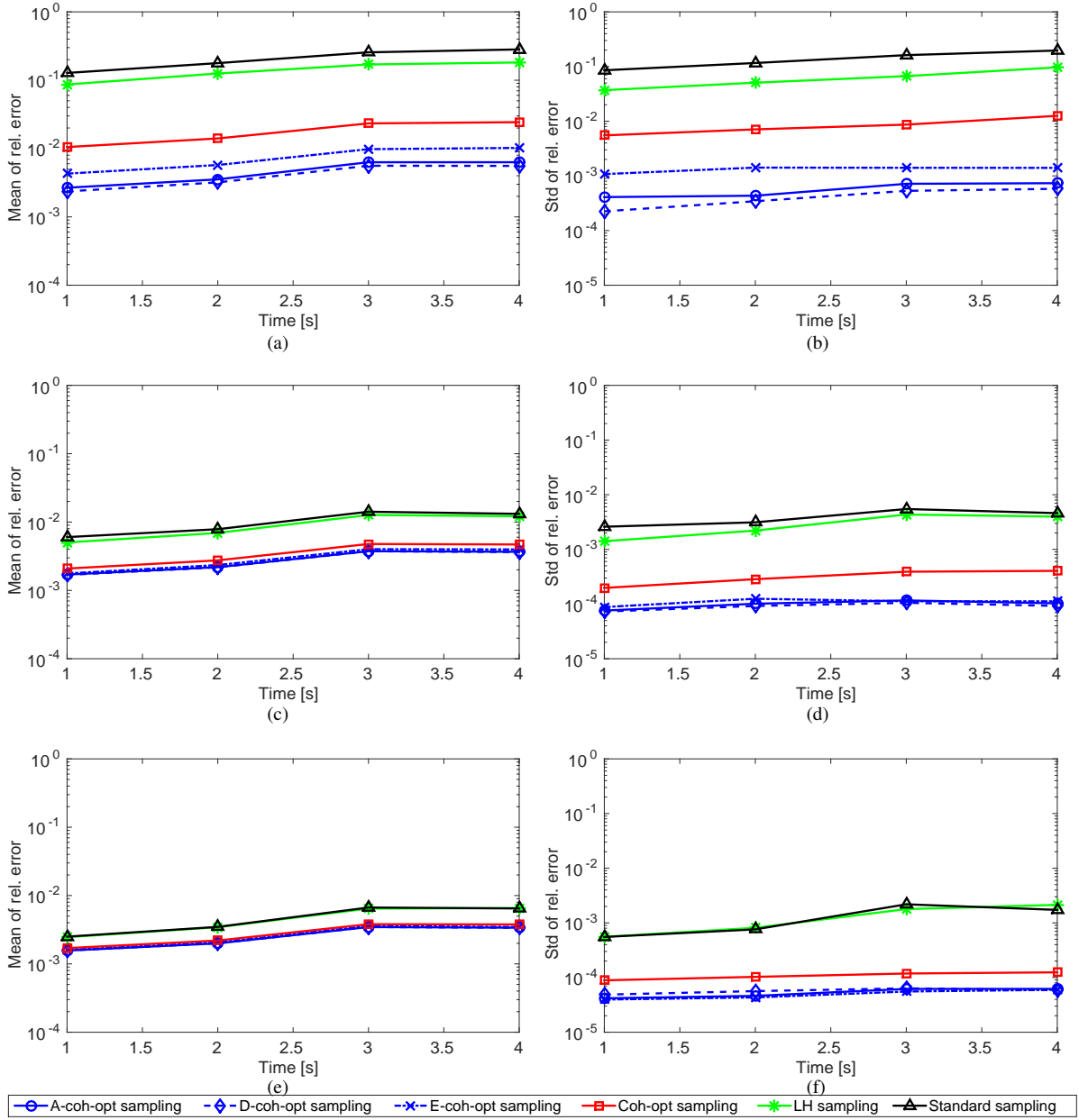
\begin{figure}
\begin{center}
{\includegraphics[width = 16.5cm]{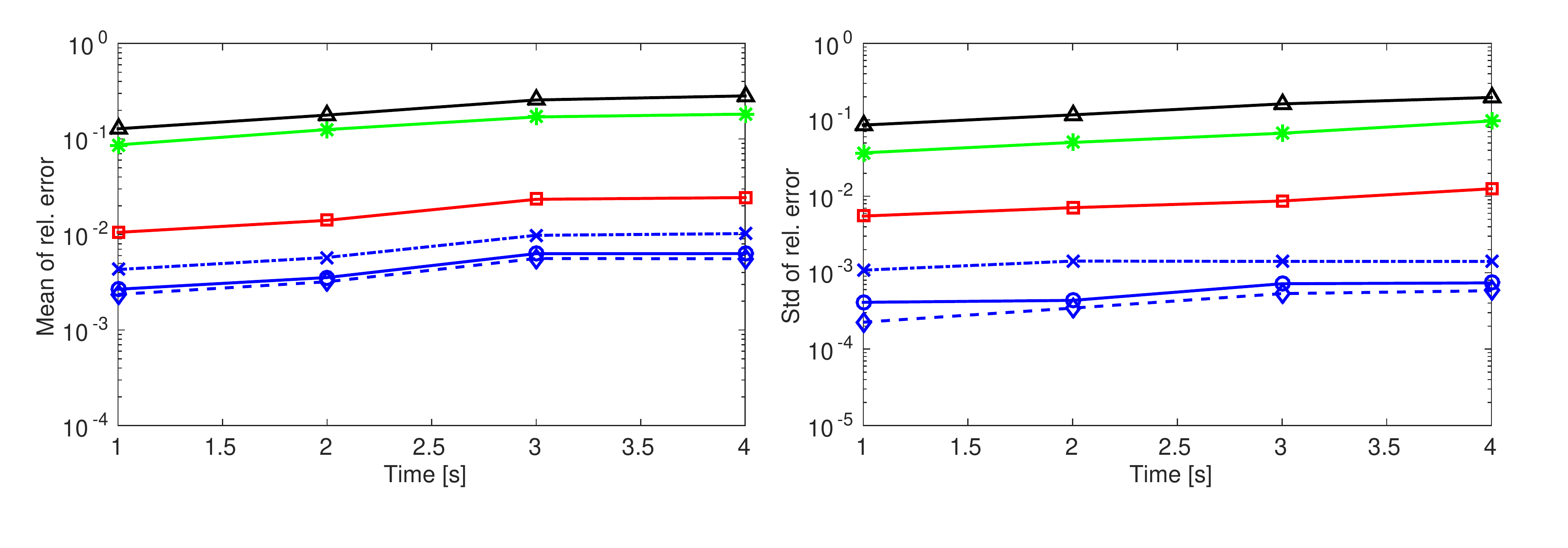}}
\put(-351,10){\text{\footnotesize (a)}}
\put(-127,10){\text{\footnotesize (b)}}
\vspace{-0.12in}
\par
{\includegraphics[width = 16.5cm]{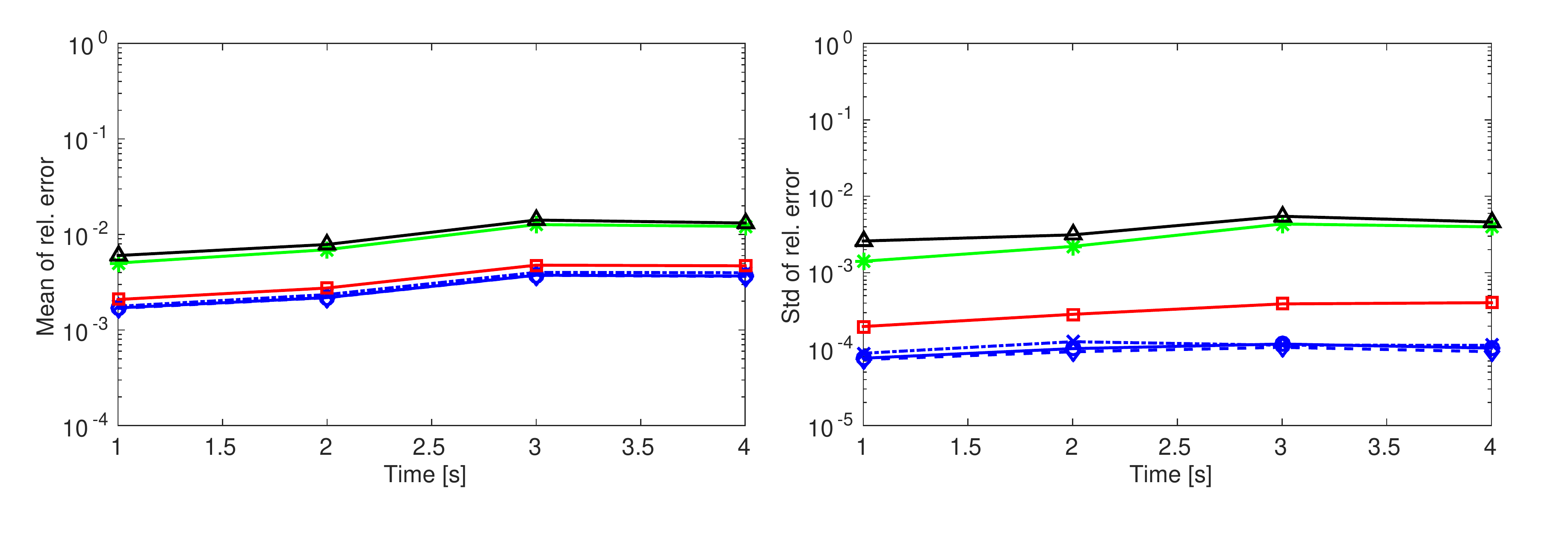}} 
\put(-351,11){\text{\footnotesize (c)}}
\put(-127,11){\text{\footnotesize (d)}}
\vspace{-0.12in}
\par
{\includegraphics[width = 16.5cm]{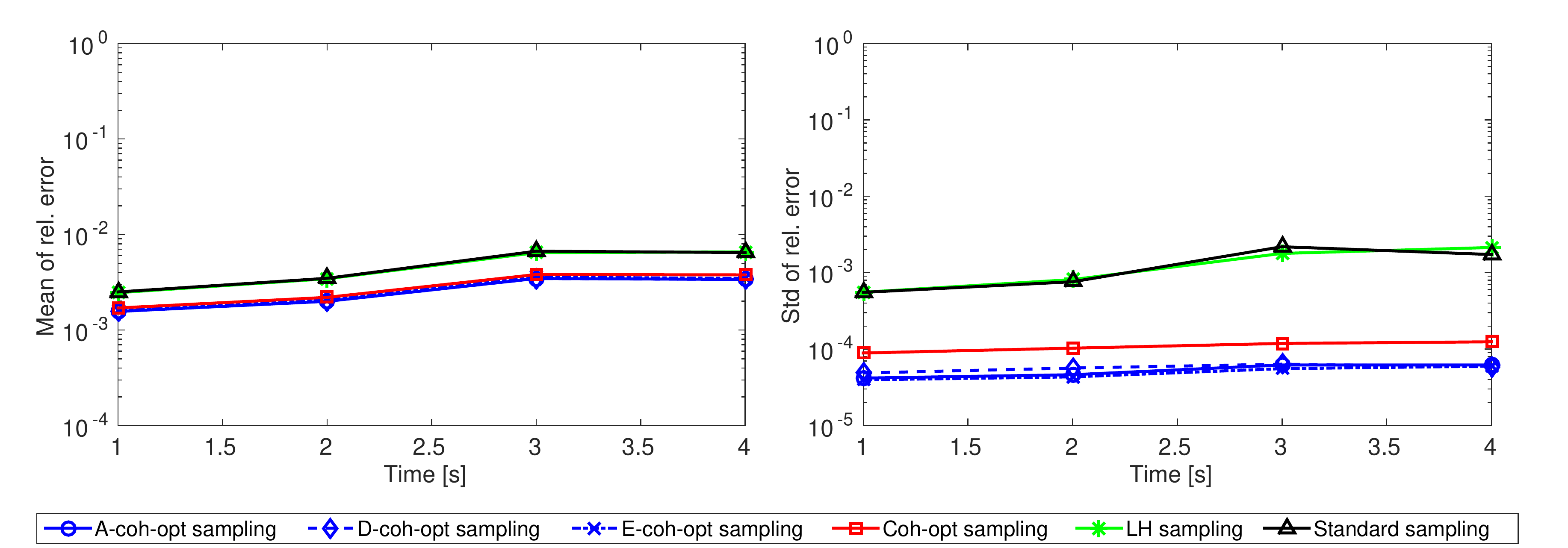}} 
\put(-351,11){\text{\footnotesize (e)}}
\put(-127,11){\text{\footnotesize (f)}}
\caption{Mean and standard deviation of the relative error in estimating the displacement $u(\bm \Xi, t)$ with a 9th order PCE and: (a)-(b): $N=242$, (c)-(d): $N=440$, (e)-(f): $N=660$.}
\label{fig:duffing_main}
\end{center}
\end{figure}

Fig. \ref{fig:duffing_main} demonstrates the mean and standard deviation of the relative error in predicting $N_v=10000$ independent realizations of the displacement obtained by different sampling methods. Three oversampling ratios of 1.1, two and three which correspond to $N=242,440$ and $660$ samples, respectively, have been considered. Following \cite{Narayan_2014}, we assume $N_c = \floor{1.5 P \log (P)}$, where $P$ is the number of unknown PC coefficients and is given in (\ref{eqn:P+1}). 

Again, the long-time integration issue we discussed earlier is observed here as well; that is, independent of the sampling technique, the relative error slightly increases over time for a fixed $p$. One possible approach to address this issue is to employ Multi-Element generalized Polynomial Chaos (ME-gPC) in which the total order of expansion is adjusted over the time. Discussion on ME-gPC is beyond the scope of this paper and we refer the interested reader to \cite{Wan05}. 

As it is shown in Fig. \ref{fig:duffing_main}, the alphabetic-coherence-optimal techniques result in smaller values for the relative error in comparison to the other methods, with { \it D-coh-opt} and { \it A-coh-opt} to perform slightly better than { \it E-coh-opt}. This superiority is signified in Figs. \ref{fig:duffing_main}a-b where the oversampling ratio is small. These plots suggest that for the case of high-order PC expansions, the alphabetic-coherence-optimal techniques are the most suitable sampling methods, in particular, when the number of samples  is slightly larger than the number of unknown coefficients $P$. For the applications where evaluation of the QoI is computationally expensive, we prefer to approximate the PCE with a minimum number of solution realizations, usually with $N$ slightly larger than $P$. 

The standard MC and LH techniques provide the least accurate results, specially, for the case of $N=242$ where the oversampling ratio is low. In this case, these two methods fail to accurately approximate the solution as the error features large mean and variance (Fig. \ref{fig:duffing_main}b). As the number of samples is increased, MC and LH methods perform considerably better such that with $N=440$ samples, both techniques result in acceptable errors. 

Similar to the previous example, the errors obtained by the coherence-optimal sampling falls somewhere between those of MC (or LH) and the alphabetic-coherence-optimal sampling methods. By comparing Figs. \ref{fig:duffing_main}a, \ref{fig:duffing_main}c, and \ref{fig:duffing_main}e, it can be seen that, as $N$ is increased, the coherence-optimal results converge to the alphabetic-coherence-optimal results at a higher rate in comparison to MC and LH methods.    

\subsection{Prediction of remaining useful life of batteries}
\label{sec:rul}
The last problem of interest in this study is the model-based estimation of the remaining useful life (RUL) of a Lithium-ion battery (LIB). RUL of an LIB is defined as the amount of time before the battery health falls below a defined threshold, e.g., reaching the cut-off potential at the end of discharge \citep{Sankararaman2015}, and is studied under the context of prognostics and health management \cite{Tang2014}. For the sake of completeness, a brief introduction to model-based prognostics of LIB's with an emphasis on RUL predictions is presented in the following. 

\subsubsection{Model-based prognostics}
\label{sec:Model_based_prognostics}

In general, prognosis approaches may be categorized as data-driven and model-based \citep{Sankavaram2009,Baraldi2013}. A mathematical representation of the system based on the physics of the problem is used in model-based methods, while data-driven techniques use system monitoring data instead \citep{Sankavaram2009}. Both techniques have been successfully employed to predict the RUL of batteries \citep{Si2011, Daigle2012,Limeng2013,Sankararaman2013b,Sankararaman14,Sankararaman2015, Liu2015}. Here, we focus on model-based RUL estimation for LIB's which is suitable for online health monitoring. 

Let us assume that the system model is given by the following set of equations, \cite{Daigle2012,Sankararaman2013b,Sankararaman14,Sankararaman2015},
\begin{eqnarray}
\label{eqn:state_space_model}
\dot{\bm{z}}(t)=\bm{f}(t, \bm{z}(t), \bm{\theta}(t), \bm{\nu}(t), \bm{v}_p(t));
\label{eqn:state_model}
 \\
\bm{y}(t)=\bm{h}(t, \bm{z}(t), \bm{\theta}(t), \bm{\nu}(t), \bm{v}_m(t)),
\label{eqn:output_model}
\end{eqnarray}
where $\bm{z}(t)$ is the state vector, $\bm{f}$ the state equation, $\bm{\theta}(t)$ the model parameter vector, $\bm{\nu}(t)$ the input vector, $\bm{v}_p(t)$ the process noise vector, $\bm{y}(t)$ the output vector, $\bm{h}$ the output equation, and $\bm{v}_m(t)$ the measurement noise vector.

As illustrated in \citep[Fig. 1]{Sankararaman2015}, the computational framework for prognostics and RUL predictions includes three main steps: estimation of the state $\bm{z}$ at time $t_p$, prediction of the future states, and finally RUL computation. Kalman or particle filtering approaches may be employed within a Bayesian framework to estimate the state vector $\bm{z}$ using the output data measured up to time $t_p$. Both Eqs. (\ref{eqn:state_space_model}) and (\ref{eqn:output_model}) should be used in the estimation step. Following \citep{Sankararaman2015}, in this study we assume that the estimated state vector $\bm{z}$ at time $t_p$ is given, and instead we focus on the prediction and RUL computation steps. 

In the prediction step, only the state space model in (\ref{eqn:state_space_model}) is used since no measured data is available for $t>t_p$, and the goal is to predict an event $E$ (depending on the states, parameters, and inputs) indicating the LIB's end of life. In the present work, $E$ occurs when the LIB's terminal voltage $V$ is smaller than a user-specified cut-off voltage $ V_{\text{cutoff}}$. More specifically, for each $t_p$, we define 
\begin{equation}
\label{eq:eod_constraints}
E(t_p) := \text{inf} \{t \in  \mathbb{R}: t\geq t_p \land V<V_{\text{cutoff}}\}
\end{equation}
and RUL as 
\begin{equation}
\label{eq:rul}
R(t_p) := E(t_p) - t_p.
\end{equation}
\subsubsection{An electrical equivalent circuit model for LIB's}
\label{sec:LIB_model_state_space}

\begin{figure}[H]
\begin{center}
{\includegraphics[width = 8.25cm]{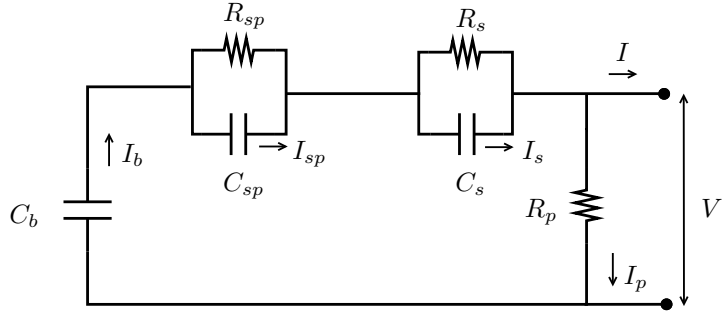}
\put(-255,33){$C_b$}
\put(-213,56){$I_b$}
\put(-175,45){$C_{sp}$}
\put(-149,57){$I_{sp}$}
\put(-175,109){$R_{sp}$}
\put(-88,45){$C_{s}$}
\put(-63,57){$I_{s}$}
\put(-88,107){$R_{s}$}
\put(-62,35){$R_p$}
\put(-25,10){$I_p$}
\put(-28,94){$I$}
\put(4,35){$V$}
}
\caption{LIB equivalent circuit model \cite{Sankararaman2013b,Sankararaman14,Sankararaman2015}.}
\label{fig:eq_circuit}
\end{center}
\end{figure}

In order to facilitate the application of the model-based RUL estimation for the online health monitoring of LIB's, the mathematical model described by Eqs. (\ref{eqn:state_space_model}) and (\ref{eqn:output_model}) is desired to be computationally inexpensive. Following closely \citep{Sankararaman2013b,Sankararaman14,Sankararaman2015}, we consider an empirical model of the LIB based on the electrical equivalent circuit model shown in Fig. \ref{fig:eq_circuit}. The charge of the LIB, denoted by $q_b$, is held by the nonlinear capacitance $C_b$. The voltage drop due to the surface over-potential is given by $R_{sp}-C_{sp}$; see Fig. \ref{fig:eq_circuit}. Additionally, the resistances $R_s$ and $R_p$ are used to model the Ohmic drop and the parasitic self-discharging, respectively, \citep{Sankararaman2013b,Sankararaman14,Sankararaman2015}. 

Denoting the LIB maximum capacity and charge by $C_{max}$ and $q_{max}$, respectively, the state of charge ($SoC$), is given by 
\begin{equation}
\label{eq:soc_state_space_model}
SoC = 1 - \frac{q_{max}-q_b}{C_{max}},
\end{equation}
 \cite{Sankararaman14}. The concentration polarization resistance $R_{sp}$ is a non-linear function of $SoC$,
\begin{equation}
\label{eq:R_CP}
R_{sp} = R_{sp0} + R_{sp1} \exp[R_{sp2} (1-SoC)], 
\end{equation}
where $R_{sp0}$, $R_{sp1}$, and $R_{sp2}$ are empirical parameters \citep{Sankararaman2013b,Sankararaman14,Sankararaman2015}. $C_b$ is also given by the nonlinear (in $SoC$) relation
\begin{equation}
\label{eq:C_b}
C_{b} = C_{b0} + C_{b1} SoC+ C_{b2} SoC^2+ C_{b3} SoC^3,
\end{equation}
with $C_{bi}$, $i=0,1,2,3$, empirically determined constants \cite{Sankararaman2013b,Sankararaman14,Sankararaman2015}. The model parameters of the considered cell are taken from \cite{Sankararaman2013b,Sankararaman14,Sankararaman2015} and shown in Table \ref{table:parameters_rul}.

\begin{table}[htb]
\tiny
\caption{LIB model parameters \cite{Sankararaman2013b,Sankararaman14,Sankararaman2015}.} 
\centering
\resizebox{6cm}{!}{%
\begin{tabular}[t]{ | >{\footnotesize } m{2cm}  | >{\footnotesize } c | }   
\hline
\textbf{Parameter} & \textbf{Value} \\ 
\hline
$R_{sp0}$ & 0.0272 $\Omega$ \\ 
\hline
$R_{sp1}$ & $1.087 \times 10^{-16}$ $\Omega$  \\ 
\hline
$R_{sp2}$ & 34.64  \\ 
\hline
$R_{s}$ & 0.0067 $\Omega$  \\ 
\hline
$R_{p}$ & 10000 $\Omega$  \\ 
\hline
$C_{b0}$ & 19.8 F \\ 
\hline
$C_{b1}$ & 1745 F \\ 
\hline
$C_{b2}$ & -1.5 F \\ 
\hline
$C_{b3}$ & -200.2 F \\ 
\hline
$C_{s}$ & 115.28 F \\ 
\hline 
$C_{sp}$ & 316.69 F \\ 
\hline 
$q_{max}$ & 31100 C \\ 
\hline 
$C_{max}$ & 30807 C \\ 
\hline 
\end{tabular}}
\label{table:parameters_rul} 
\end{table} 

Voltage drops across the individual circuit elements and their associated currents and charges are governed by the set of equations presented in Table \ref{table:equi_circuit}, where $I$ denotes the discharge current. The terminal voltage of the cell $V$ is given by $V=V_b - V_{sp} - V_s$. We assume that in this example $V_{\text{cutoff}} = 16 \text{V}$.

\begin{table}[htb]

\caption{Equations describing the electrical equivalent circuit model for LIB's \cite{Sankararaman2013b,Sankararaman14,Sankararaman2015}.} 
\centering
\resizebox{13cm}{!}{%
\begin{tabular}[t]{ | >{\footnotesize} p{2cm}  | >{\footnotesize} c | }   
\hline 
  & \textbf{Governing equation}  \\   
\hline 
\centering
\textbf{Voltages} & 
{ \footnotesize $V_b = q_b/C_b, \quad V_{sp} = q_{sp}/C_{sp}, \quad V_s = q_s/C_s, \quad V_p = V_b - V_{sp} - V_s $ } \\ 
\hline
\centering
\textbf{Currents} & 
{ \footnotesize $ I_p = V_p/R_p, \quad I_b = I_p + I, \quad I_{sp} = I_b -  V_{sp}/R_{sp}, \quad I_s = I_b -  V_{s}/R_{s} $}  \\ 
\hline
\centering
\textbf{Charges} & 
{ \footnotesize $ \dot{q_b} =-I_b, \quad \dot{q}_{sp} =I_{sp}, \quad  \dot{q_{s}} =I_{s} $ }\\
\hline
\end{tabular}}
\label{table:equi_circuit} 
\end{table}

Different uncertainty sources such as those in the future inputs, state, model and process noise have been identified for the stochastic RUL estimations \citep{Sankavaram2009, Daigle2012, Saha2012,Daigle2013,Sankararaman2013b,Sankararaman14,Sankararaman2015}. We refer the readers to the provided references for more details on the sources of uncertainty for this problem. In this example, we assume that the battery is discharged at a constant current (in Amps) represented by a random variable following a Beta distribution with shape parameters $\alpha = 21.2$ and $\beta = 31.8$. As mentioned earlier, we assume that the state estimates are already obtained. For this example, we consider an estimation of the state variables, $q_b$, $q_{sp}$, and $q_s$, given in \citep[Figs. 5-7]{Sankararaman2015}, respectively, with a coefficient of variation (COV) of 0.1. Whenever the estimated state is equal to zero, we assume its standard deviation is equal to 0.1. In addition, three process noise terms with zero mean normal distributions and variances of 0.1, $10^{-4}$, and $10^{-6}$ are considered in the state equations describing $q_b$, $q_{sp}$, and $q_s$, respectively. Assuming that there is no model uncertainty, the total stochastic dimension of this problem is therefore $d=7$. We note that the univariate PC basis functions $\psi_{{j}_k}(\Xi_k)$ for this problem consist of Jacobi and Hermite polynomials which correspond to Beta and normal random variables, respectively.  

\begin{figure}[H]
\begin{center}
{\includegraphics[width = 16.5cm]{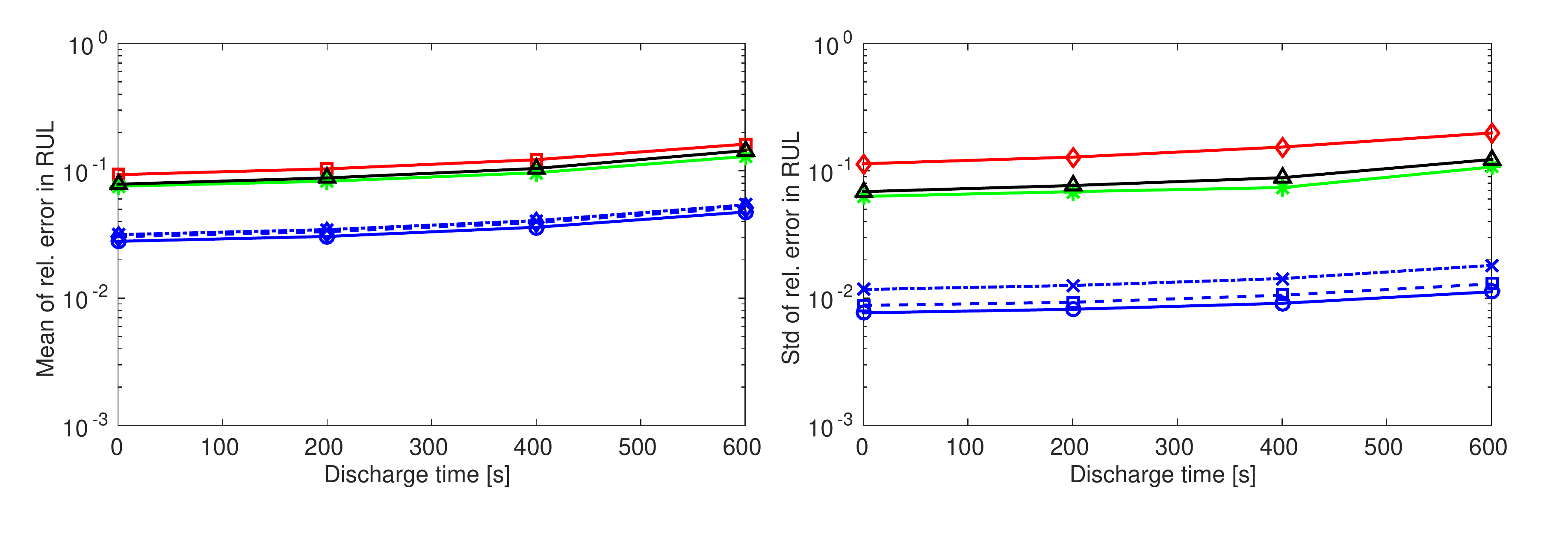}}
\put(-361,10){\text{\footnotesize (a)}}
\put(-127,10){\text{\footnotesize (b)}}
\vspace{-0.12in}
\par
{\includegraphics[width = 16.5cm]{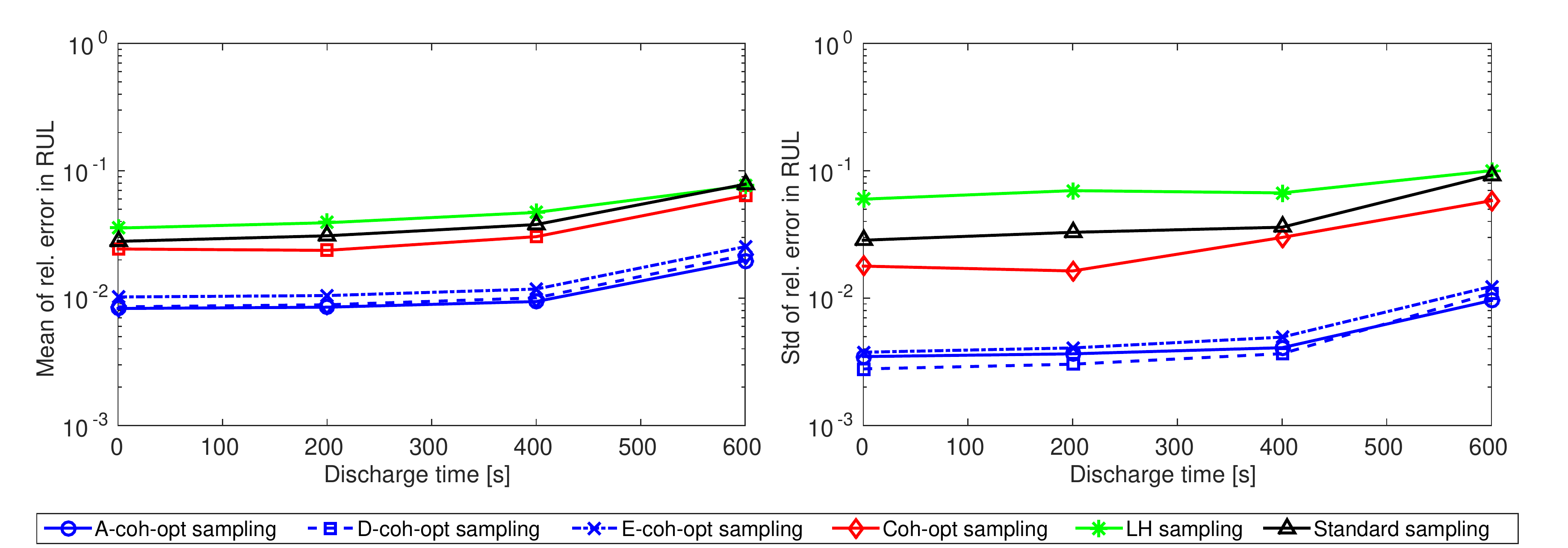}} 
\put(-361,11){\text{\footnotesize (c)}}
\put(-127,11){\text{\footnotesize (d)}}
\caption{Mean and standard deviation of the relative error in estimating the RUL; (a)-(b): $p=2$, $P=36$, $N=37$, (c)-(d): $p=3$, $P=120$, $N=121$.}
\label{fig:V_SOC_mean_sd_er}
\end{center}
\end{figure}

Fig. \ref{fig:V_SOC_mean_sd_er} shows the relative error of the sampling techniques in estimating the RUL of LIB at $t=0, 200, 400$, and 600 seconds. Similar to the previous example, 10000 independent samples are used to compute the relative error. Since model-based RUL predictions are mainly used in online applications where fast computations are critical, we limit the number of sample points in this example to be $N=P+1$, which corresponds to 36 and 121 samples for the second and third order PCE's, respectively. Similar to the previous example, we assumed $N_c = \floor{1.5 P \log (P)}$. Fig. \ref{fig:V_SOC_mean_sd_er} demonstrates that similar to the high-order PC expansions, alphabetic-coherence-optimal techniques perform better that other methods for the high-dimensional problems as well when the oversampling ratio is low. Moreover, it can be seen that for the case of $d > p$ with low oversampling ratios, coherence-optimal method is not necessarily advantageous over the standard MC and LH samplings. 

\begin{figure}[ht]
\begin{center}
{\includegraphics[width = 8.25cm]{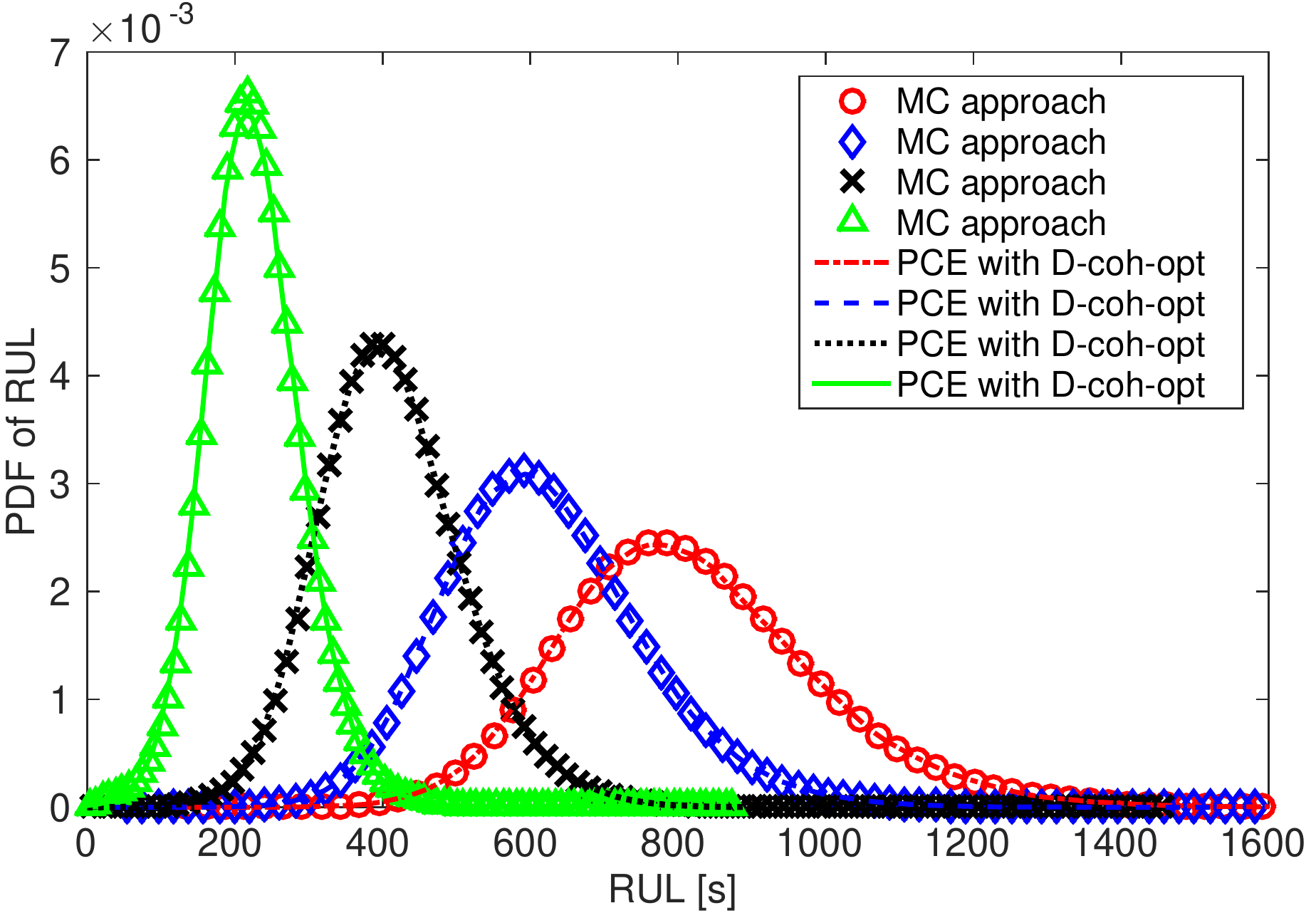}}
\caption{pdfs of the estimated RUL at $t=0, 200, 400, 600$ seconds demonstrated in red, blue, black and green colors, respectively. {\it D-coh-opt} solution is obtained with a 3rd order PCE with $N=121$ samples, while $20 \times 10^6$ samples are used to obtain the MC-based pdf.}
\label{fig:RUL_PDF}
\end{center}
\end{figure}

It is shown in Fig. \ref{fig:V_SOC_mean_sd_er}c-d that the {\it D-coh-opt} approach results in the smallest errors. We next compare the pdf of the RUL predicted with a $p=3$ PCE, obtained using the {\it D-coh-opt} approach, against a MC-based pdf in Fig. \ref{fig:RUL_PDF}. Although the pdf's obtained by these two different approaches coincide, we note that $20 \times 10^6$ solution realizations are used in constructing the MC pdf, while {\it D-coh-opt}-based PCE only requires 121 samples. This highlights the efficacy of the PCE approach, equipped with optimal sampling methods, for the online RUL prediction of LIB's. 

\section{Conclusion and summary}
\label{sec:Conclusion}

We provided an extensive review of the sampling techniques available in the literature for PCE's via least squares approximation. We also proposed a hybrid sampling approach referred to as {\it alphabetic-coherence-optimal} in which, given the same sample size, a combination of alphabetic-optimal designs, such as \textit{D}-, \textit{A}-, and \textit{E}-optimal, and the recently developed coherence-optimal technique results in more accurate approximations of the PC coefficients in comparison to other sampling methods. 

In order to provide insight for practitioners seeking the best sampling method in building PCE's via least squares approximation, we compared the empirical performance of several sampling techniques through their application to three problems involving uniform, Gaussian, and Beta random inputs, and for both low and high order (and dimension) expansions. The highlights of our observations, backed up by theoretical results, are summarized below:

\begin{itemize}
\item High-order PCE with $p>d$: We observed that for problems in which high-order expansions are required to meet accuracy requirements, the alphabetic-coherence-optimal techniques deliver the most accurate approximations. Among the alphabetic-coherence-optimal techniques, {\it D-coh-opt} results were slightly better than those obtained by {\it A-coh-opt}. In our experiments, we observed that the construction of the {\it D-coh-opt} sample sets is considerably faster than that of the {\it A-coh-opt} samples; hence, we recommend {\it D-coh-opt} for high-order PCE's. We note that generating input samples according to the alphabetic-coherence-optimal design is more expensive than sampling the inputs, for instance, randomly according to their joint pdf. However, in most practical engineering applications, generating realizations of the output QoI is considerably expensive, so the extra computational cost for sampling the inputs may be ignored. 

\item High-dimensional PCE with $d>p$: Although the alphabetic-coherence-optimal techniques still perform better than other methods considered here for the case of $d>p$, the standard MC and LH sampling methods may also deliver accurate approximations when low-order expansions are sufficient. In such cases, one may employ standard MC, LH, or other advanced space-filling sampling techniques discussed in this study as they may be less computationally expensive to generate than the alphabetic-coherence-optimal approach. 

\item The oversampling ratio $N/P$: In the numerical examples provided in this study, it was observed that when $N$ is 3-4 times larger than the number of PC coefficients $P$, one can expect that the standard MC and LH sampling methods exhibit a fairly acceptable performance as long as the order $p$ of the expansion is small. When high order expansions are required, even with a fairly large oversampling ratio, these methods may fail to provide an accurate PC approximation. For the cases with small values of $N/P$, alphabetic-coherence-optimal techniques outperform other methods considerably, specially when high-order PCE's are employed. 

\end{itemize}

Finally, we note that existing algorithms for solving alphabetic-optimal design problems are computationally expensive for large values of $P$, thus limiting their applicability to relatively low-order and/or low-dimensional PCE's. Developing scalable optimization techniques to address this difficulty remains an open problem in ODE. 

\section*{Acknowledgements}

MH's work was supported by the National Science Foundation grant CMMI-1201207. This material is based upon work of AD supported by the U.S. Department of Energy Office of Science, Office of Advanced Scientific Computing Research, under Award Number DE-SC0006402 and NSF grant CMMI-1454601. AD is thankful of Prof. Paul Constantine from Colorado School of Mines for fruitful discussions on least squares approximation and regression.

\clearpage

\appendix
\appendixpage
\addtocontents{toc}{\protect\setcounter{tocdepth}{0}}

\section{Derivation of bounds in Corollary \ref{cor:stability}}
\label{apx:proof_cor}
Using the assumption $\Vert \bm M- \bm I\Vert\le \delta$, $0\le \delta<1$, and the symmetry of $\bm M- \bm I$, we have that 
\[
\Vert \bm M - \bm I \Vert = \vert\lambda_{\max}(\bm M - \bm I)\vert\le \delta.
\]
This implies $\vert\lambda(\bm M - \bm I)\vert\le \delta$, which in turn gives $\lambda_{\max}(\bm M)-1\le \delta$ and $-\delta\le \lambda_{\min}(\bm M) - 1$; hence, the inequality in (\ref{eqn:stab_bound1}). A direct application of (\ref{eqn:stab_bound1}) gives (\ref{eqn:stab_bound2}). 

To see (\ref{eqn:stab_bound3}), we first note that $\Vert\hat{u}\Vert_{L_2(\Omega,f)}^2= \mathbb{E}[\hat{u}^2]=\Vert\hat{\bm{c}}\Vert^2$ due to the orthonormality of the PC basis. From (\ref{eqn:regression2}) and following \cite{Cohen2013}, we have
\[
\Vert\hat{\bm{c}}\Vert \le \Vert \left((\bm{W}  \bm{\Psi})^T (\bm{W} \bm{\Psi})\right)^{-1}\Vert \cdot\Vert \left(\bm{W} \bm{\Psi}\right)^T\Vert\cdot \Vert \bm{W} \bm{u}\Vert.  
\]
To bound $\Vert \left(\bm{W} \bm{\Psi}\right)^T\Vert$, we first observe that $\Vert \left(\bm{W} \bm{\Psi}\right)^T\Vert = \Vert\bm{W} \bm{\Psi}\Vert = \Vert(\bm{W}  \bm{\Psi})^T (\bm{W} \bm{\Psi})\Vert^{1/2}$. Notice that $\Vert 1/N(\bm{W}  \bm{\Psi})^T (\bm{W} \bm{\Psi})\Vert\le\Vert 1/N(\bm{W}  \bm{\Psi})^T (\bm{W} \bm{\Psi})-\bm I\Vert +1\le(1+\delta)$. Therefore,
\begin{equation}
\label{eqn:stab_bound_first}
\Vert \left(\bm{W} \bm{\Psi}\right)^T\Vert\le \sqrt{N}(1+\delta)^{1/2}.
\end{equation}
To bound $\Vert \left((\bm{W}  \bm{\Psi})^T (\bm{W} \bm{\Psi})\right)^{-1}\Vert$, we follow a similar analysis as in \cite[Section 7.2]{Rauhut10}. Specifically, 
\[
\Vert N\left((\bm{W}  \bm{\Psi})^T (\bm{W} \bm{\Psi})\right)^{-1}\Vert = \Vert \sum_{k=0}^{\infty}\left(\bm I - \bm M\right)^k\Vert\le \sum_{k=0}^{\infty} \Vert \bm I - \bm M\Vert^k\le \sum_{k=0}^{\infty} \delta^k=\frac{1}{1-\delta}, 
\]
which implies
\begin{equation}
\label{eqn:stab_bound_second}
\Vert \left((\bm{W}  \bm{\Psi})^T (\bm{W} \bm{\Psi})\right)^{-1}\Vert\le \frac{1}{N}\frac{1}{1-\delta}.
\end{equation}
Putting together (\ref{eqn:stab_bound_first}) and (\ref{eqn:stab_bound_second}), we arrive at the inequality (\ref{eqn:stab_bound2}).

Notice that (\ref{eqn:perturb_bound}) is a standard result for which we refer the reader to \cite[Section 3.5]{Hansen12}. 
\section{Matrix determinant update formula}
\label{apx:det_update}

Let $\bm A$ be an invertible matrix of size $P \times P$, and let $\bm b$ and $\bm a$ be two vectors of size $P$. Then
\begin{equation}
\label{eqn:update_det}
\frac{|\bm A \pm \bm b \bm a^T|}{|\bm A|}=1 \pm \bm a^{T} {\bm A}^{-1}  \bm b. 
\end{equation}
\begin{proof}
When $\bm A = \bm I$, we have
\begin{equation}
\left[\begin{array}{cc} \bm{I} & \bm 0 \\ \bm a^{T} & 1 \end{array}\right] \left[\begin{array}{cc} \bm{I} \pm \bm b \bm a^T & \bm b \\ \bm 0 & 1 \end{array}\right]\left[\begin{array}{cc} \bm{I} & \bm 0 \\ \mp \bm a^{T} & 1 \end{array}\right] = \left[\begin{array}{cc} \bm{I} & \bm b \\ \bm 0 & \bm{I} \pm  \bm a^T \bm b \end{array}\right]. \nonumber
\end{equation}
Taking the determinant of both sides results in $|\bm I \pm \bm b \bm a^T| = 1 \pm \bm a^T \bm b$. Finally, one may write $|\bm A \pm \bm b \bm a^T|=|\bm A (\bm I \pm \bm A^{-1} \bm y \bm a^T)|=|\bm A|(1 \pm \bm a^{T} {\bm A}^{-1}  \bm b)$.
\end{proof}

\section{Matrix trace update formula}
\label{apx:trace_update}

Let $\bm A$ be an invertible matrix of size $P \times P$, and let $\bm b$ and $\bm a$ be two vectors of size $P$. Additionally, assume $1 \pm \bm a^T \bm{A}^{-1} \bm b \ne 0$. Then
\begin{equation}
\label{eqn:update_trace}
\text{Tr} \left( (\bm A \pm \bm b \bm a^T)^{-1}\right)= \text{Tr} \left(  \bm A^{-1} \right) \mp \text{Tr} \left( \frac{\bm A^{-1} \bm b \bm a^T \bm A^{-1}}{1 \pm \bm a^T \bm{A}^{-1} \bm b} \right).
\end{equation}
\begin{proof} 
Sherman-Morrison rank-one update formula states
\begin{equation}
\label{eqn:Sherman-Morrison}
(\bm A \pm \bm b \bm a^T)^{-1}= \bm A^{-1} \mp \left( \frac{\bm A^{-1} \bm b \bm a^T \bm A^{-1}}{1 \pm \bm a^T \bm{A}^{-1} \bm b}\right),
\end{equation}
which can be easily verified by substituting (\ref{eqn:Sherman-Morrison}) in identity $(\bm A \pm \bm b \bm a^T)(\bm A \pm \bm b \bm a^T)^{-1}=\bm I$. Taking the trace of both sides of (\ref{eqn:Sherman-Morrison}) completes the proof. 
\end{proof}

\clearpage

\section*{References}
{\footnotesize
\bibliographystyle{siam}
\bibliography{Bib_V1}}

\end{document}